\documentclass[structabstract]{aa}  
\usepackage{graphicx}
\usepackage{txfonts}
\usepackage{subfigure}
\usepackage{longtable}
\usepackage{booktabs}
\usepackage{natbib}
\usepackage{ulem}

\def \figwidth {8cm}
\def \deg {^\circ}
\def \kmps {\mathrm{km}\mathrm{s}^{-1}}
\def \radpd {\mathrm{rad}\,\mathrm{d}^{-1}}
\def \qfrac {\frac{q_2}{q_1}}
\def \sini {\sin{i}}
\def \vsini {v\sini}
\def \alphasini {\frac{\alpha}{\sqrt{\sini}}}
\def \alphaimax {\alpha_{90\deg}}
\def \alphaimin {\alpha_{10\deg}}
\def \Omegasini {\Omega{\sqrt{\sini}}}
\def \Omegaimax {\Omega_{90\deg}}
\def \Psini {\frac{P}{\sini}}
\def \aap {A\&A}
\def \Teff {T_\mathrm{eff}}
\def \etal {et~al.}
\def \Msun {M_\odot}

\def \BarnesV {2005MNRAS.357L...1B}
\def \Eggleton{2008MNRAS.389..869E}
\def \Gizon {2004SoPh..220..169G}
\def \Gray {1989ApJ...341..421G}
\def \Flower {1996ApJ...469..355F}
\def \Kaufer {1999Msngr..95....8K}
\def \Kawaler {1999ApJ...516..349K}
\def \Kenyon {1995ApJS..101..117K}
\def \KuekerV {2005A&A...433.1023K}
\def \KuekerVII {2007AN....328.1050K}
\def \Kitchatinov {1999A&A...344..911K}

\def \ReinersIIa {2002A&A...384..155R}
\def \ReinersIIb {2002A&A...388.1120R}
\def \ReinersIIIa {2003A&A...398..647R}
\def \ReinersIIIb {2003A&A...412..813R}
\def \ReinersIIIc {2003A&A...408..707R}
\def \ReinersIV {2004A&A...415..325R}

\def \ReinersVI {2006A&A...446..267R}
\def \Press {1992nrca.book.....P}
\def \Schou {1998ApJ...505..390S}
\def \Siess {SDF00}
\def \Suarez {2007CoAst.150..211S}

\begin{document}
   \title{New measurements of rotation and differential rotation in A-F stars: Are there two populations of differentially rotating stars? \thanks{Based on observations made with ESO Telescopes at the La Silla Paranal Observatory under programme ID's 074.D-0008 and 075.D-0340.}\thanks{Tables~\ref{tab:results} and \ref{tab:comp_diffrot} are available online only.}}


   \author{Ammler-von Eiff, M.
          \inst{1,2,3}
          \and
          Reiners, A.\inst{2}
          }

   \institute{Th\"uringer Landessternwarte Tautenburg, Sternwarte 5, 07778 Tautenburg, Germany\\
             \email{ammler@tls-tautenburg.de}
\and
Institut f\"ur Astrophysik, Georg-August-Universit\"at G\"ottingen, Friedrich-Hund-Platz 1, 37077 G\"ottingen, Germany\\
              \email{areiners@astro.physik.uni-goettingen.de}
         \and
             Max Planck Institute for Solar System Research, Max-Planck-Stra{\ss}e 2, 37191 Katlenburg-Lindau, Germany
                     }

   \date{Received September 15, 1996; accepted March 16, 1997}

 
  \abstract
   {The Sun displays differential rotation that is intimately connected to the solar dynamo and hence related to solar activity, the solar cycle, and the solar wind. Considering the detectability and habitability of planets around other stars it is important to understand the role of differential rotation in other stars.
   }
   {We present projected rotational velocities and new measurements of the rotational profile of some 180 nearby stars with spectral types A-F. The results are consolidated by a homogeneous compilation of basic stellar data from photometry and the identification of multiple stellar systems. New and previous measurements of rotation by line profile analysis are compiled and made available.
   }
   {The overall broadening profile is derived analysing spectral line shape from hundreds of spectral lines by the method of least-squares deconvolution, reducing spectral noise to a minimum. The effect of differential rotation on the broadening profile is best measured in inverse wavelength space by the first two zeros of its Fourier transform. 
   }
   {Projected rotational velocity $\vsini$ is measured for more than 110 of the sample stars. Rigid and differential rotation can be distinguished in 56 cases where $\vsini>12\,\kmps$. We detect differential rotation rates of $\frac{\delta\Omega}{\Omega}=5\,\%$ and more. Ten stars with significant differential rotation rates are identified. The line shapes of 43 stars are consistent with rigid rotation, even though differential rotation at very low rates might still be possible in these cases. The strongest amount of relative differential rotation (54\,\%) detected by line profile analysis is found among F stars. 
}
   {As of now, 33 differential rotators detected by line profile analysis have been confirmed. The frequency of differential rotators decreases towards high effective temperature and rapid rotation. There is evidence for two populations of differential rotators, one of rapidly rotating A stars at the granulation boundary with strong horizontal shear and one of mid- to late-F type stars with moderate rates of rotation and less shear. The gap in between can only partly be explained by an upper bound found for the horizontal shear of F stars. Apparently, the physical conditions change at early-F spectral types. The range of horizontal shear observed for mid-type F stars is reproduced by theoretical calculations while there seems to be a discrepancy in period dependence for late-F stars.}


   \maketitle
%

\section{Introduction}

The Sun does not rotate like a rigid body. Instead, angular rotational velocity varies with latitude (latitudinal differential rotation) and with distance from the center (radial differential rotation). 
Differential rotation is thought to emerge from the interaction between the turbulent motions in the convective envelope and Coriolis forces which are due to rotation. It is believed to be an important ingredient of the widely accepted solar $\alpha$-$\Omega$ dynamo\footnote{The parameter $\alpha$ of the dynamo must not be confused with the parameter $\alpha$ of differential rotation!}. Thus, differential rotation is closely connected to magnetic activity.

While latitudinal differential rotation can be observed directly on the solar surface, e.g. by tracing solar spots, radial differential rotation in the Sun is assessed indirectly by helioseismology \citep[e.g.][]{\Schou}. The amount of latitudinal surface differential rotation is expressed by angular velocity $\Omega$ as function of latitude $l$. The solar differential rotation is well described by the surface rotation law 
\begin{equation}
\label{eq:rotlaw}
{\Omega}(l)=\Omega_\mathrm{Equator}(1-\alpha\,{\sin}^2{l})
\end{equation}
with the parameter $\alpha=0.2$ of relative differential rotation. While relative differential rotation $\alpha$ is the quantity measured in the present work, it is the horizontal shear ${\Delta\Omega}$ which connects the measurements to the description of stellar physics:
\begin{equation}
\label{eq:shear}
{\Delta\Omega}=\Omega_{\mathrm{Equ}}-\Omega_{\mathrm{Pol}}=\alpha\Omega_{\mathrm{Equ}}
\end{equation}

The potential to assess stellar differential rotation by asteroseismology has been studied e.g. for white dwarfs \citep{\Kawaler}, solar-type stars \citep{\Gizon}, and B stars \citep{\Suarez} using rotationally split frequencies of stellar oscillations.

Stellar differential rotation cannot be observed directly since the surface of stars can usually not be imaged with sufficient resolution. Yet, one may expect active stars to be rotating differentially assuming that Sun-like dynamos also work on other stars. On the other hand, a magnetic field may also inhibit differential rotation through Lorentz forces so that it is not clear what degree of differential rotation to expect in active stars.

In the case of the solar rotation pattern, the angular momentum transport by the gas flow in the convective envelope offers an explanation. \citet{\Kitchatinov}, \citet{\KuekerV}, and \citet{\KuekerVII} studied Sun-like differential rotation in main-sequence and giant stars based on mean-field hydrodynamics.
According to these models, horizontal shear varies with spectral type and rotational period and is expected to increase strongly towards earlier spectral type while there is a weaker variation with rotational period. Horizontal shear vanishes for very short and very long periods and becomes highest at a rotational period of 10\,days for spectral type F8 and 25\,days in the case of a model of the Sun. A strong period dependence is also seen in models of stars with earlier spectral type. \citet{\KuekerVII} reproduce observed values of horizontal shear of mid-F type stars of more than $1\radpd$.

Although stars are unresolved, differential rotation can be assessed by indirect techniques through photometry and spectroscopy. 
Photometric techniques use the imprints of stellar spots in the stellar light curve. If the star rotates differentially, spots at different latitudes will cause different photometric periods which are detected in the light curve and allow one to derive the amount of differential rotation.
Stellar spots further leave marks in rotationally-broadened absorption lines of stellar spectra. These features change their wavelength position when the spot is moving towards or receding from the observer due to stellar rotation. Then, Doppler imaging recovers the distribution of surface inhomogeneities by observing the star at different rotational phases. Using results from Doppler imaging, \citet{\BarnesV} studied differential rotation of rapidly rotating late-type stars and confirmed that horizontal shear strongly increases with effective temperature.

Both photometry and Doppler imaging need to follow stellar spots with time and thus are time-consuming. Another method, line profile analysis, takes advantage of the rotational line profile which has a different shape in the presence of differential rotation. While rigid rotation produces line profiles with an almost elliptical shape, solar-like differential rotation results in cuspy line shapes.

In contrast to the methods mentioned above, line profile analysis does not need to follow the rotational modulation of spots. Instead, only one spectrum is sufficient. Therefore, hundreds of stars can be efficiently checked for the presence of differential rotation. Line profile analysis is complementary to other methods in the sense that it works only for non-spotted stars with symmetric line profiles.

The line profile analysis used in the present work was introduced by \citet{\ReinersIIa}. The noise of the stellar line profile virtually vanishes by least squares deconvolution of a whole spectral range averaging the profiles of hundreds of absorption lines. The Fourier transform of the rotational profile displays characteristic zeros whose positions are characteristic of the rotational law. Thus, the amount of differential rotation can be easily measured in Fourier space. In principle, the influence of all broadeners like turbulent broadening, instrumental profile and rotational profile simply is a multiplication in Fourier space. Therefore, turbulent broadening and instrumental profile leave the zeros of the rotational profile unaffected.

Among hundreds of stars with spectral type A to early K, 31 differential rotators have so far been identified by line profile analysis \citep{\ReinersVI}. Actually, many more differential rotators might be present which escaped detection because of low projected rotational velocity $\vsini$, low amounts of differential rotation below the detection limit, shallow or strongly blended line profiles, asymmetric line profiles, or multiplicity. 

The amount of differential rotation depends on spectral type and $\vsini$. \citet{\ReinersVI} note that the frequency of differential rotators as well as the amount of differential rotation tends to decrease towards earlier spectral types. Convective shear seems strongest close to the convective boundary. While many slow Sun-like rotators ($\vsini\approx10\,$km/s) with substantial differential rotation are found, only few rapid rotators and hot stars are known to display differential rotation. 

The main goal of the present work is to better understand the frequency and strength of differential rotation throughout the HR diagram, in particular the rapid rotators and the hotter stars close to the convective boundary. New measurements are combined with previous assessments of line profile analysis and presented in a catalogue containing the measurements of rotational velocities $\vsini$, characteristics of the broadening profile, differential rotation parameters $\alpha$, and pertinent photometric data (Sect.~\ref{sect:results}). Furthermore, stellar rotational velocities and differential rotation is studied in the HR diagram (Sect.~\ref{sect:HRD}) and compared to theoretical predictions (Sect.~\ref{sect:shear}).

\section{Observations}
\label{sect:obs}
The techniques applied in the present work are based on high-resolution spectroscopy with high signal-to-noise ratio. Therefore, this work is restricted to the brightest stars. In order to improve the number statistics of differential rotators throughout the HR diagram, the sample of the present work completes the work of previous years \citep{\ReinersIIIa,\ReinersIIIb,\ReinersIV,\ReinersVI} and consists of 182 stars of the southern sky known to rotate faster than $\approx10\,$km/s, to be brighter than $V=6$, and with colours $0.3 < (B-V) < 0.9$. Emphasis was given to suitable targets rather than achieving a complete or unbiased sample.

The spectra were obtained with the FEROS and CES spectrographs at ESO, La Silla (Chile). Five stars were observed with both FEROS and CES. The data was reduced in the same way as by \citet{\ReinersIIIa,\ReinersIIIb}.

Spectra of 158 stars studied in the present work were taken with FEROS at the ESO/MPG-2.2m telescope between October 2004 and March 2005 (ESO programme 074.D-0008). The resolving power of FEROS \citep{\Kaufer} is fixed to 48,000 covering the visual wavelength range (3600{\AA}-9200{\AA}).	
Spectra of 24 objects with $\vsini\lesssim45\,$km/s were obtained with the CES spectrograph at the ESO-3.6m telescope, May 6-7, 2005 (ESO programme 075.D-0340). A spectral resolving power of 220,000 was used.


\section{Methods}

\subsection{Least-squares deconvolution}
The reduced spectra are deconvolved by least squares deconvolution of a wide spectral region following the procedure of \citet{\ReinersIIa}. 

The wavelength region 5440-5880{\AA} has already been used successfully by \citet{\ReinersIIIb} for FEROS spectra of FGK stars and is thus used in the present work, too. In the case of A stars, the number of useful lines in this range is insufficient and one has to move to the blue spectral range. \citet{\ReinersIV} analyzed A stars in the range $4210-4500$\,{\AA} using spectra taken with ECHELEC at ESO, La Silla. This wavelength range inconveniently includes the H$\gamma$ line right in the centre. In the present work, we take advantage of the wide wavelength coverage of FEROS and find that the range $4400-4800\,${\AA} between H$\gamma$ and H$\beta$ contains a sufficient number of useful spectral lines. The CES spectra span a clearly narrower wavelength range of about 40\,{\AA}. \citet{\ReinersIIIa} succeeded with CES spectra in the ranges 5770-5810\,{\AA} and $6225-6270$\,{\AA}. Within the present work the full range covered by our CES spectra of $6140-6175$\,{\AA} is used.

The deconvolution process derives the average line profile from a multitude of observed line profiles. The spectral regions used with the FEROS and CES spectra display a sufficient number of useful absorption lines and are unaffected by strong features and telluric bands. Although the deconvolution algorithm successfully disentangles lines blended by rotation, the deblending degenerates if lines are too close to each other. However, the deconvolution was found to be robust in the wavelength ranges used. The regions can be used homogeneously for all the stars analyzed \citep{\ReinersIIIb}.

A template is generated as input to the processing, containing information about the approximate strength and wavelength positions of spectral lines. The line list is drawn from the Vienna Astrophysical Line Data-Base \citep[VALD;][]{Kupka00,Kupka99,Ryabchikova97,Piskunov95} based on a temperature estimate derived from spectral type. VALD is queried with the {\it extract stellar} feature and a detection limit of 0.02.

The optimization of the line profile (by regularised least square fitting of each pixel of the line profile) alternates with the optimization of the equivalent widths (by Levenberg-Marquardt minimization -- see \citealp{\Press} -- of the difference of observed and template equivalent widths). Only fast rotators with $\vsini\gtrsim40\,$km/s are accessible with the FEROS spectra. Then, the rotational profile dominates all other broadening agents. In the case of the CES spectra, much narrower profiles are traced at $\vsini$ as low as $\approx10\,$km/s. Therefore, these spectra are deconvolved using \underline{P}hysical \underline{L}east \underline{S}quares \underline{D}econvolution ({\em PLSD}, \citealp{\ReinersIIIa}), accounting for the different thermal broadening of the spectral lines of the involved atomic species.

\subsection{Analysis of the overall broadening profile in Fourier space}
\label{sect:fourier}
The shape of the deconvolved rotational profile is characteristic of the type of rotation. Although differential rotation may be distinguished in wavelength space, the analysis works best in the frequency domain \citep{\ReinersIIa}. The Fourier transform of the rotational profile displays characteristic zeros. The ratio of the first two zeros depends on the amount of differential rotation. The convolution with other broadening agents, e.g. the instrumental profile, corresponds to a multiplication in Fourier space and thus reproduces the zeros. The projected rotational speed has to be sufficiently high to resolve the frequency of the second zero. The Nyquist frequency of {\em FEROS} spectra with a resolving power of 48,000 is 0.08\,s/km and for CES it is 0.37\,s/km. In order to properly trace the broadening profile to the maximum of the second side lobe, $\vsini$ has to be at least 45\,$\kmps$ (FEROS) and 12\,$\kmps$ (CES), respectively \citep{\ReinersIIIa,\ReinersIIIb}. Actually, zero positions can be measured for even lower $\vsini$ but then they are beyond the actual sampling limit and affected by noise features. Then, an interpretation is not possible.

\subsection{Derivation of the projected rotational velocity $\vsini$}
The projected rotational velocity is taken as the average of the values derived from each one of the two zero positions in the Fourier transform. If the second zero is no longer resolved reliably (according to the $\vsini$ limits given above), $\vsini$ is obtained from the first zero only.

The formal error bar does not reflect systematic uncertainties which correspond to a relative error of $\approx5\%$ \citep{\ReinersIIIb}. In order to obtain a conservative error estimate, the error of 5\,\% is adopted. The formal error is taken if it is larger than 5\,\%.

\subsection{Derivation of the parameter of differential rotation $\alpha$}
\label{sect:alpha}
The parameter $\alpha$ of differential rotation is calculated from the ratio of the zeros of the Fourier transform $\qfrac$ using the modeled relation of $\qfrac$ and $\alpha$ given in \citet[][fig.~2]{\ReinersIIa}. Unknown inclination represents the largest uncertainty when interpreting measurements of rotational broadening from line profile analysis. In order to reflect this uncertainty, the parameter $\alpha$ of differential rotation was calculated in the present work for inclination angles $10\deg$ and $90\deg$. Period is assessed accordingly, assuming inclinations of $10\deg$ and $90\deg$. Additional uncertainties originate from the unknown limb darkening. Values of the differential rotation parameter are also derived from the calibrations given in \citet{\ReinersIIIa} for comparison and consistency with previous work. There, the function $\alphasini$ is obtained which minimizes the scatter due to unknown inclination.

A detection limit of $\alpha=5\,\%$ originates from the unknown limb darkening and the consequent inability to pin down differential rotation rates lower than $\approx5\,\%$. In the following, all stars with $\alpha\gtrsim5\,\%$ are considered Sun-like differential rotators in agreement with \citet{\ReinersVI}. Adopting a Sun-like rotation law (Eq.~\ref{eq:rotlaw}), the maximum possible value of Sun-like differential rotation is $\alpha=100\,\%$ implying that the difference of the rotation rates of the poles and the equator equals the angular velocity at the equator. Negative values of $\alpha$ would imply anti-solar differential rotation with angular velocities being higher at higher latitudes. However, negative values of $\alpha$ are not interpreted in terms of true differential rotation. Rather it may be understood as rigid rotation in the presence of a cool polar spot mimicking differential rotation \citep{\ReinersIIb}. In principle, the cuspy line profile of solar-like differential rotation might also be mimicked by spots, namely in the configuration of a belt of cool spots around the equator. However, this is a hypothetic case which has never been observed.

There are further effects causing spurious signatures of differential rotation. In the case of very rapid rotation ($\vsini>200\kmps$), the spherical shape of the stellar surface is distorted by centrifugal forces and the resulting temperature variations modify the surface flux distribution. This so-called gravitational darkening occurs in particular in the case of rapidly-rotating A and F stars \citep{\ReinersIIIc,\ReinersIV}. In the case of close binaries, the spherical shape of the star can be modified by tidal elongation and thus change the surface flux distribution.

It is actually the absolute horizontal shear $\Delta\Omega$ which is related to the stellar interior physics while it is only the relative value $\alpha=\frac{\Delta\Omega}{\Omega}=\frac{\Delta{\Omega}P}{2\pi}$ that can be measured by line profile analysis. 
Rotational periods are derived from photometric radius and measured $\vsini$, assuming inclinations of $10\deg$ and $90\deg$ to be consistent with the values of $\alpha$ used. Thus, values of horizontal shear are used which were obtained for both inclination angles to account for uncertainty about inclination.
Because of the relation to $\alpha$, very low values of horizontal shear may thus be detected in the case of very slow rotators which however is limited by the minimum $\vsini$ required for line profile analysis. In the case of very rapid rotators, however, differential rotators might be undetectable even at strong horizontal shear.

\section{Peculiar line shapes and multiplicity}
\label{sect:peculiar}
The deconvolved profiles of many stars show peculiarities like asymmetries, spots, or multiple components. We widely treat these profiles according to \citet{\ReinersIIIa}. Instead of skipping all stars with peculiar profiles, we tried to get as much information as possible from the line profile, tagging the results with a flag which indicates the type of the peculiarity.

Many profiles are asymmetric in that the shape of the blue wing of the overall profile is different from the red wing. In these cases, the analysis is repeated for each of the wings separately. The average of the results is adopted with conservative error bars.

In the case of distortions caused by multiples, three cases are distinguished. In the first case, the components are separated well and can be analyzed each. In the second case, the profile is blended but dominated by one component. If the secondary component can be fully identified it is removed or a line wing unaffected by the secondary is taken. Third, if the components cannot be disentangled, then only an upper limit on $\vsini$ is derived from the overall broadening profile.

In some cases, the deconvolved profile is symmetric, even though the star is a known or suspected multiple. This may affect both the assessment of global stellar properties and the interpretation of line profiles:
\begin{itemize}
\item In the case of two very similar but spatially unresolved components, the colours and effective temperature will be the same as for a single object. However, luminosity will be multiplied by the number of components. This results in a considerable shift in the HR diagram and thus to a wrong assessment of luminosity class (and ${\log}g$).
\item In the case of very different, unresolved components, the derived parameters will be dominated by the primary, i.e. the brightest component, but a considerable shift in the HR diagram is still possible.
\item In favourable cases, spatially unresolved components show up as separate spectral components and can be separated for the rotational analysis -- as was discussed above. In the worst-case scenario, there is no relative shift between the spectral components forming an apparently single symmetric profile. This might be the case if components of a spectroscopic multiple are almost on the line of sight at the time of observation or if the components form a wide but spatially unresolved multiple involving slow orbital motion.
\end{itemize}

Therefore, in order to detect analyses which might be affected by multiplicity, the catalogue of \citet{\Eggleton} was searched. Suspicious analyses are identified by one of two flags. There is one 'photometric' flag 'x' and one 'spectroscopic' flag 'y'. The photometric data is considered spurious and flagged by 'x' if the star has spatially unresolved components. In some cases, the spectral types of the components are similar but no individual brightness measurements are given in the literature. In such a case, we derive an estimate of the primary magnitude by applying an offset of +0\fm75 to the given total magnitude of the system. If the star is a known or suspected spectroscopic multiple, results from apparent single-component line profiles might be spurious and are thus flagged by 'y'. The analysis also gets the 'y' flag if there are known visual components closer than $3\farcs0$ and at a brightness difference of less than $5\,$mag in the $V$ band. These values are based on experience with slit and fibre spectrographs under typical seeing conditions without adaptive optics.

\section{Discussion of measurements}
\label{sect:results}
\label{sect:basic}

Results for all stars of the sample are presented in Table~\ref{tab:results}. For clarity, only one peculiarity flag is given with the results. Our choice is that flags indicating multiplicity override flags denoting asymmetries or bumps. In particular, spectral components of multiples might also be affected by asymmetries. Therefore, the reader is advised to check the other flag, too, which indicates the type of the measurement.

\onllongtab{1}{
\begin{longtable}{lrrrrrrrrrl}
\caption{\label{tab:results}Results obtained in the present work from the analysis of the broadening profile and pertinent stellar data. The first 25 data rows provide measurements from CES spectra while the remainder lists the results from FEROS spectra. The last column provides flags which characterize the broadening profile and the type of measurement done for a particular object: {\bf d} differential rotator, {\bf r} consistent with $\Delta\Omega=0$, i.e. rigid rotation, {\bf p} anti-solar differential rotation (probably cool polar cap), {\bf v} only measurement of $\vsini$, \textbf{$<$} only upper limit on $\vsini$, {\bf g} symmetric profile, {\bf a} asymmetric profile (e.g. spots and multiplicity), {\bf f} noisy, distorted profile, failed deconvolution, {\bf m2} spectroscopic binary, {\bf m3} spectroscopic triple, {\bf m?} multiple with unknown number of components. For clarity, only one flag is given to indicate peculiarity. See the text for more details. Two additional flags refer to indications of multiplicity found in the literature: {\bf x} photometry affected by unresolved components, and {\bf y} spectroscopic analysis affected by unresolved components.}\\
\hline\hline
\multicolumn{2}{l}{name}&$(B-V)$&$T_\mathrm{eff}$&$M_\mathrm{V}$&$\vsini$&$\qfrac$&$\alphasini$&$\Psini$&$\Delta\Omegasini$&flags\\
\multicolumn{2}{l}{}&&[K]&&[kms$^{-1}$]&&&$\mathrm{d}$&$[\radpd]$&\\
\hline
\endfirsthead
\multicolumn{11}{l}{\it Table \ref{tab:results} continued}\\
\hline\hline
\multicolumn{2}{l}{name}&$(B-V)$&$	T_\mathrm{eff}$&$M_\mathrm{V}$&$\vsini$&$\qfrac$&$\alphasini$&$\Psini$&$\Delta\Omegasini$&flags\\
\multicolumn{2}{c}{}&&[K]&&[kms$^{-1}$]&&&$\mathrm{d}$&$[\radpd]$&\\
\hline
\endhead
\hline
\multicolumn{11}{r}{\it Table \ref{tab:results} continued next page}\\
\endfoot
\hline
\multicolumn{11}{r}{\it End of Table \ref{tab:results}}\\
\hline
\endlastfoot
HD32743             &     & 0.42& 6632& 3.28& 22.7$\,\pm\,$  1.1&1.62$\,\pm\,$0.02& 0.19$\,\pm\,$ 0.09&  3.3& 0.36$\,\pm\,$ 0.17&dg        \\         
HD33262             &     & 0.53& 6158& 4.38& 15.4$\,\pm\,$  0.8&1.68$\,\pm\,$0.01& 0.10$\,\pm\,$ 0.07&  3.5& 0.18$\,\pm\,$ 0.13&dg        \\         
HD64379             &     & 0.47& 6409& 3.88& 43.1$\,\pm\,$  2.2&1.72$\,\pm\,$0.01& 0.03$\,\pm\,$ 0.06&  1.4& 0.13$\,\pm\,$ 0.27&ra        \\         
HD68456             &     & 0.44& 6541& 3.19&  8.8$\,\pm\,$  1.1&2.25$\,\pm\,$0.01&     &  9.2&     &vfx       \\                                     
HD80671             &     & 0.41& 6678& 3.13&  7.4$\,\pm\,$  1.0&1.34$\,\pm\,$0.00&     & 10.7&     &$<$m2x      \\                                   
HD89569             &     & 0.48& 6366& 3.03& 11.2$\,\pm\,$  0.6&1.81$\,\pm\,$0.00&     &  8.2&     &vf        \\                                     
HD91324             &     & 0.50& 6282& 3.19&  8.3$\,\pm\,$  0.4&1.93$\,\pm\,$0.14&     & 10.7&     &vf        \\                                     
HD104731            &     & 0.42& 6632& 3.23& 15.9$\,\pm\,$  3.2&1.17$\,\pm\,$0.02& 0.54$\,\pm\,$ 0.11&  4.8& 0.70$\,\pm\,$ 0.14&dg        \\
\end{longtable}
}

\onllongtab{2}{
\begin{longtable}{lrrrrrrrll}
\caption{\label{tab:comp_diffrot}Compilation of all stars measured to date by line profile analysis and pertinent stellar data. Line profile measurements ($\vsini$, $\qfrac$, $\alphasini$, $\Omegasini$, and flags) are adopted from the source indicated in the last column: {\bf 1} this work (CES measurements), {\bf 2} this work (FEROS measurements), {\bf 3} \citet{\ReinersIIIa}, {\bf 4} \citet{\ReinersIIIb}, {\bf 5} \citet{\ReinersIV}, {\bf 6} \citet[table B1]{\ReinersVI}, {\bf 7} \citet[table B2]{\ReinersVI}. The basic stellar data ($\Teff$, $M_\mathrm{V}$, and $\Psini$) are only partly taken from the given sources as is described in more detail in the main body of this paper. Flags are given as in Table~\ref{tab:results} for the present work which widely follows the use of flags in previous work. In \citet{\ReinersIIIb}, there are additional flags {\bf b} for suspected blended multiples, {\bf s} for suspected spotted stars, and {\bf c} for stars with symmetric profile but which are candidates for spottedness.}\\
\hline\hline
name&$T_\mathrm{eff}$&$M_\mathrm{V}$&$\vsini$&$\qfrac$&$\alphasini$&$\frac{P}{\sini}$&$\Delta\Omegasini$&flags&source\\
&[K]&&[kms$^{-1}$]&&&$[\mathrm{d}]$&$[\radpd]$&\\
\hline
\endfirsthead
\multicolumn{10}{l}{\it Table \ref{tab:comp_diffrot} continued}\\
\hline\hline
name&$	T_\mathrm{eff}$&$M_\mathrm{V}$&$\vsini$&$\qfrac$&$\alphasini$&$\frac{P}{\sini}$&$\Delta\Omegasini$&flags&source\\
&[K]&&[kms$^{-1}$]&&&$[\mathrm{d}]$&$[\radpd]$&\\
\hline
\endhead
\hline
\multicolumn{10}{r}{\it Table \ref{tab:comp_diffrot} continued next page}\\
\endfoot
\hline
\multicolumn{10}{r}{\it End of Table \ref{tab:comp_diffrot}}\\
\hline
\endlastfoot
HD    432                 & 6763& 0.96&  71.0$\,\pm\,$  3.6& 1.78$\,\pm\,$ 0.03& 0.00$\,\pm\,$ 0.04&   3.0&     &          & 6\\                                                    
HD    493                 & 6678& 2.05& 116.8$\,\pm\,$  5.8& 1.68$\,\pm\,$ 0.00& 0.10$\,\pm\,$ 0.06&   1.1& 0.56$\,\pm\,$ 0.34&dgy       & 2\\                                      
HD    693                 & 6366& 3.51&   3.8$\,\pm\,$  0.8&     &     &  19.4&     &          & 3\\                                                                                
HD    739                 & 6496& 3.55&   2.4$\,\pm\,$  1.2&     &     &  28.9&     &$<$         & 3\\                                                                              
HD   1581                 & 6002& 4.56&   2.3$\,\pm\,$  0.4&     &     &  22.7&     &$<$         & 3\\                                                                              
HD   2151                 & 5855& 3.45&   3.3$\,\pm\,$  0.3&     &     &  28.1&     &          & 3\\                                                                                
HD   2726                 & 6916& 2.48&  13.5$\,\pm\,$  1.5&     &     &   7.3&     &x         & 3\\                                                                                
HD   3302                 & 6453& 2.72&  17.8$\,\pm\,$  0.8&     &     &   5.8&     &          & 3\\
\end{longtable}
}
\begin{figure}
\includegraphics[width=\figwidth]{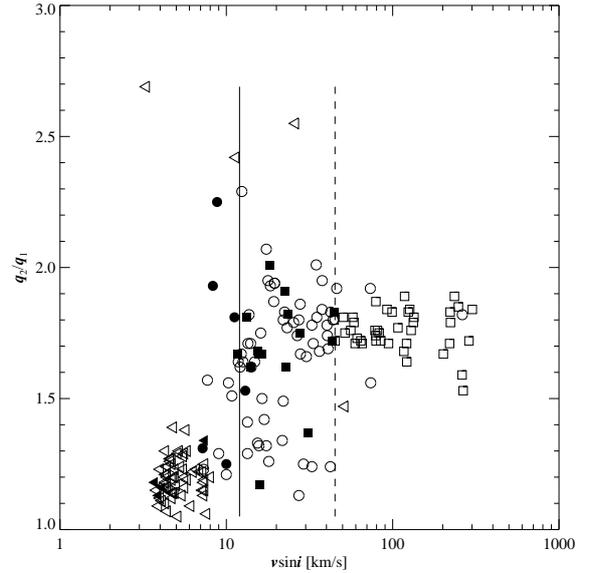}
\caption{\label{fig:vsini} Measurements of $\vsini$ and $\qfrac$ of the present work. Open symbols indicate measurement from FEROS spectra and filled symbols data obtained from CES spectra. Squares show measurements of fully resolved rotational profiles. Circles denote the cases where only $\vsini$ was measured reliably. The triangles show measurements of upper limits to rotational velocities. The vertical lines indicate the sampling limits of CES (solid) and FEROS (dashed).}
\end{figure}

Figure~\ref{fig:vsini} displays the measurements of $\qfrac$ and $\vsini$ obtained in the present work. $\vsini$ measurements range from $4\,\kmps$ to $\approx300\,\kmps$ while assessments of $\qfrac$ can be as low as $\approx\,1$ and reach values of up to 2.7. As is discussed in Sect.~\ref{sect:fourier}, the measurements of $\qfrac$ are indicative of the mode of rotation only at $\vsini\gtrsim12\,\kmps$ (CES) and $\vsini\gtrsim45\,\kmps$ (FEROS), in total 56 objects. If one zero of the Fourier transform is beyond the sampling limit, $\vsini$ can still be measured. In total, $\vsini$ was measured for 114 objects. If both zeros of the Fourier transform are beyond the sampling limits of CES and FEROS, resp., the $\vsini$ derived must be interpreted as an upper limit. Indeed, these values represent the lowest measured, are not much below $4\,\kmps$, and cluster at the lowest measured velocities. The few upper limits at $\qfrac\approx2.5$ are all due to peculiar line profiles caused by binarity or spots. Upper limits on $\vsini$ were derived in 68 cases.

\subsection{Measurements of multiple systems}
\label{sect:mult}
Among the sample, there are 21 stars with indications of multiplicity detected in the overall broadening profile. Although multiplicity complicates the line profile analysis severely, a number of interesting objects could be measured.

One major achievement of the present work is the analysis of resolved profiles of spectroscopic binaries. Among the 21 stars in the sample with line profiles indicative of multiplicity, rigid rotation could be assessed for \object{HD\,4150}, \object{HD\,17168}, and \object{HD\,19319}.

In addition, there are nine spectroscopic binaries with both spectroscopic components analysed: \object{HD\,16754}, \object{HD\,26612}, \object{HD\,41547}, \object{HD\,54118}, \object{HD\,64185}, \object{HD\,105151}, \object{HD\,105211}, \object{HD\,147787}, and \object{HD\,155555}. The line wings could be analysed separately and the results are shown in Table~\ref{tab:results}.

One of these objects, HD\,64185, displays signatures of differential rotation. It is a binary system, possibly even triple. Two spectroscopic components are detected in the present work and analysed separately. The stronger broad component (considered \object{HD\,64185\,A} in the present work) indicates differential rotation with the strongest absolute horizontal shear encountered in the present work's sample while the weaker and narrow component (\object{HD\,64185\,B}) is consistent with rigid rotation. However, as HD\,64185 is possibly a triple, it cannot be excluded that the line profile of HD\,64185\,A is a blend mimicking the signatures of differential rotation. Individual stellar parameters of the components are not available so that the position in the HR diagram might be erroneous. Therefore, the measured relative differential rotation and horizontal shear are excluded from the studies in the present work. Nevertheless, additional studies of this star are encouraged since HD\,64185\,A shows the strongest horizontal shear among the differentially rotating F-type stars.

Among the nine spectroscopic binaries with both components measured, HD\,147787 and HD\,155555 are particularly interesting since they have been studied by line profile analysis before and are analysed once more in the present work. This means that line profiles are compared obtained at different orbital phase. HD\,147787 was found to be asymmetric by \citet{\ReinersIIIb}. The present work resolved the profile based on a CES spectrum and resolved two spectroscopic components for which $\vsini$ is measured. HD\,147787 is a spectroscopic binary with a period of 40 days according to \citet{2008MNRAS.389..869E}. HD\,155555 did not stand out particularly in the work of \citet{\ReinersIIIa} but the present work detects two spectroscopic components. It is a short-period pre-main sequence spectroscopic binary according to \citet[][\object{V824 Arae}]{2000A&A...360.1019S}. They also derived $\vsini$ of both components which agree with the values assessed in the present work within the error bars. \citet{2008MNRAS.387.1525D} measured significant rates of differential rotation for both components from Doppler imaging. As these translate to amounts of relative differential rotation of a few percent only, they are below the detection limit of line profile analysis. In the present work, the quality of the deconvolved profiles is insufficient and affected by asymmetries, so that the rotational profile could not be measured anyway, even though the rotational velocities are sufficiently large. The asymmetries probably are due to the spottedness which on the other side enabled the Doppler imaging analyses by \citet{2008MNRAS.387.1525D}.

We also mention \object{HD\,80671} which does not display resolved spectroscopic components but has been studied before. It is a blended multiple suspected by \citet{\ReinersIIIb}. In fact, it is a spectroscopic binary with a close visual companion at a separation of $0\farcs066$ and a period of 3 years \citep{2000AJ....119.3084H,2008MNRAS.389..869E}. Based on CES spectroscopy, the present work detects two blended spectroscopic components. Only an upper limit on $\vsini$ of the primary is derived.

It is equally important to also discuss known multiples with an apparent single line profile. Signatures of differential rotation have to be regarded with care if the star does not show resolved spectroscopic components although it is a known multiple. The catalogue of \citet{\Eggleton} was searched systematically for multiples among the stars of the sample. Such evidence indicates the need of further studies but is incapable of firmly rejecting candidates of differential rotation by itself. The findings in the work of \citet{\Eggleton} are discussed in the following in what concerns stars with signatures of differential rotation:
\begin{itemize}
\item \object{HD\,493} is a visual binary with similar components and a separation of 1\farcs433. It cannot be excluded that the light of both components entered the slit of the spectrograph which would produce a composite spectrum. The spectrum does not indicate any spectroscopic multiplicity but the line profile could still be composed of two components. Additional studies of this object are encouraged since it is located right in the lack of differential rotators in the HR diagram at early-F spectral types (Sect.~\ref{sect:populations}).
\item  Signatures of differential rotation were detected for \object{HD\,105452} by \citet{\ReinersIIIa}. This cannot be reproduced in the present work, the spectrum displays an asymmetric profile instead. HD\,105452 is a known spectroscopic binary. Possibly, the blended line profile mimicked differential rotation in older spectra at an orbital phase when there is no displacement of spectral lines. Another explanation could be that the emitted flux is modified by tidal elongation depending on orbital phase (see Sect.~\ref{sect:alpha}).
\item \object{HD\,114642} is single according to \citet{\Eggleton} but displays an asymmetric line profile in the present work. The line wings are analysed separately and both wings indicate differential rotation.
\item \object{HD\,124425} is a spectroscopic binary with a short period of $2.696\,$d.  Although, it appears single in the spectrum available to this work, one cannot exclude that the broadening profile is the sum of two blended profiles or affected by tidal elongation. The star deserves particular attention since the line profile indicates the second largest amount of differential rotation ($\alpha=44\,\%$) among the F stars of the sample.
\end{itemize}
We also mention measurements of rigid rotators which might be affected by multiplicity. There is one group of stars, \object{HD\,15318}, \object{HD\,42301}, and \object{HD\,57167}, which appear single in the spectrum but are listed as binaries with unknown component parameters by \citet{\Eggleton}. This is also the case for \object{HD\,215789} which shows signatures expected for anti-solar differential rotation or a polar cap.

The multiples are excluded from the studies of relative differential rotation and horizontal shear presented in Sect.~\ref{sect:HRD} and \ref{sect:shear} since their position in the HR diagram and the line profile measurements might be erroneous. This also concerns differential rotators detected in previous work. These stars are: \object{HD\,17094}, \object{HD\,17206}, \object{HD\,18256}, \object{HD\,90089}, \object{HD\,121370}, and \object{HD\,182640}. Nevertheless, the data might still be correct but this needs to be proven by further studies. The measurements obtained for these stars are tabulated in Table~\ref{tab:results}. These data flagged with 'm', 'x', or 'y' offer a starting point for future studies.

\subsection{Comparison with previous work and discussion of systematic uncertainties}
\label{sect:uncertainties}
\begin{figure}
\subfigure[\label{fig:vsini_resid}]{\includegraphics[width=\figwidth]{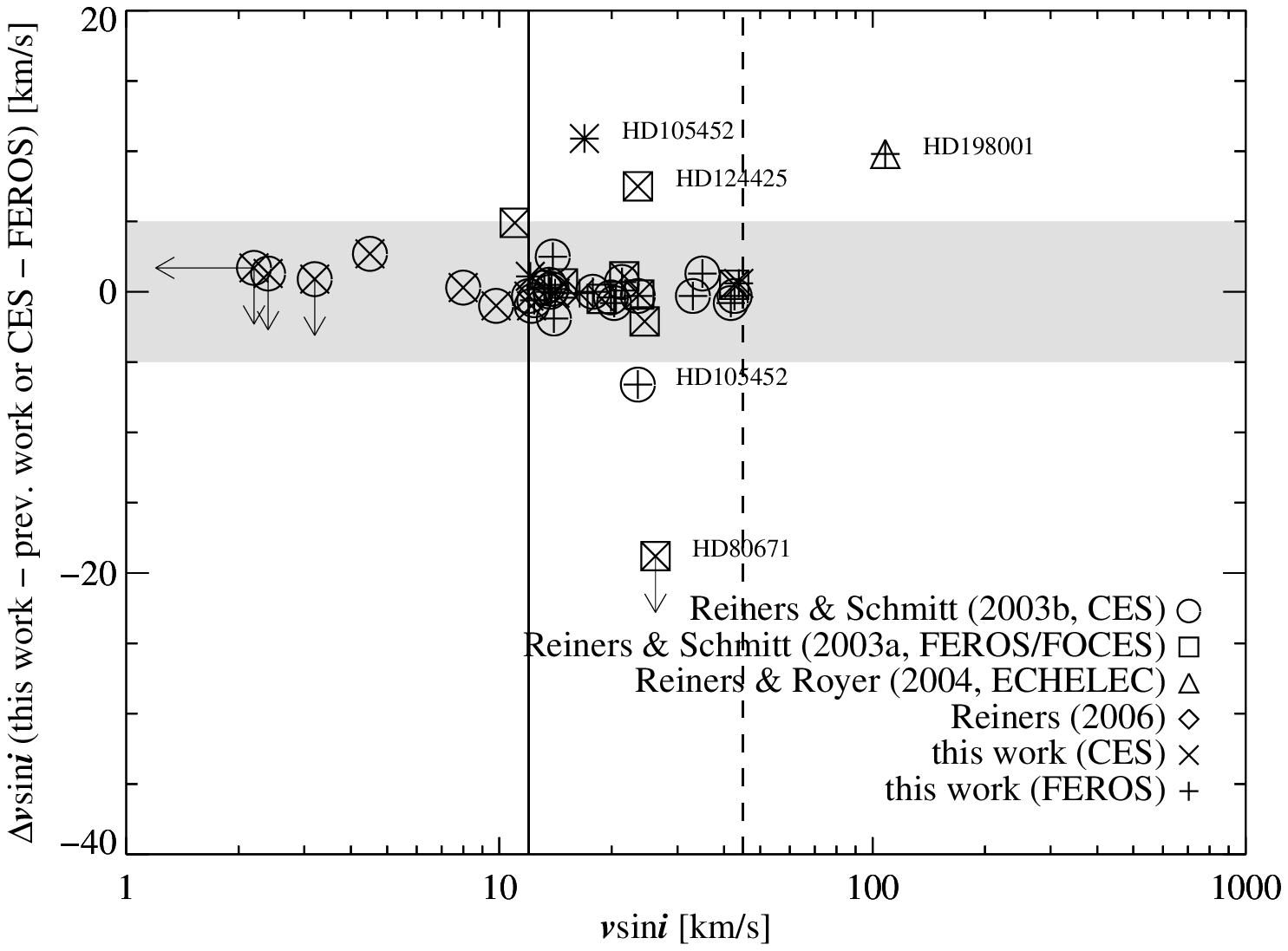}}
\subfigure[\label{fig:q2q1_resid_vsini}]{\includegraphics[width=\figwidth]{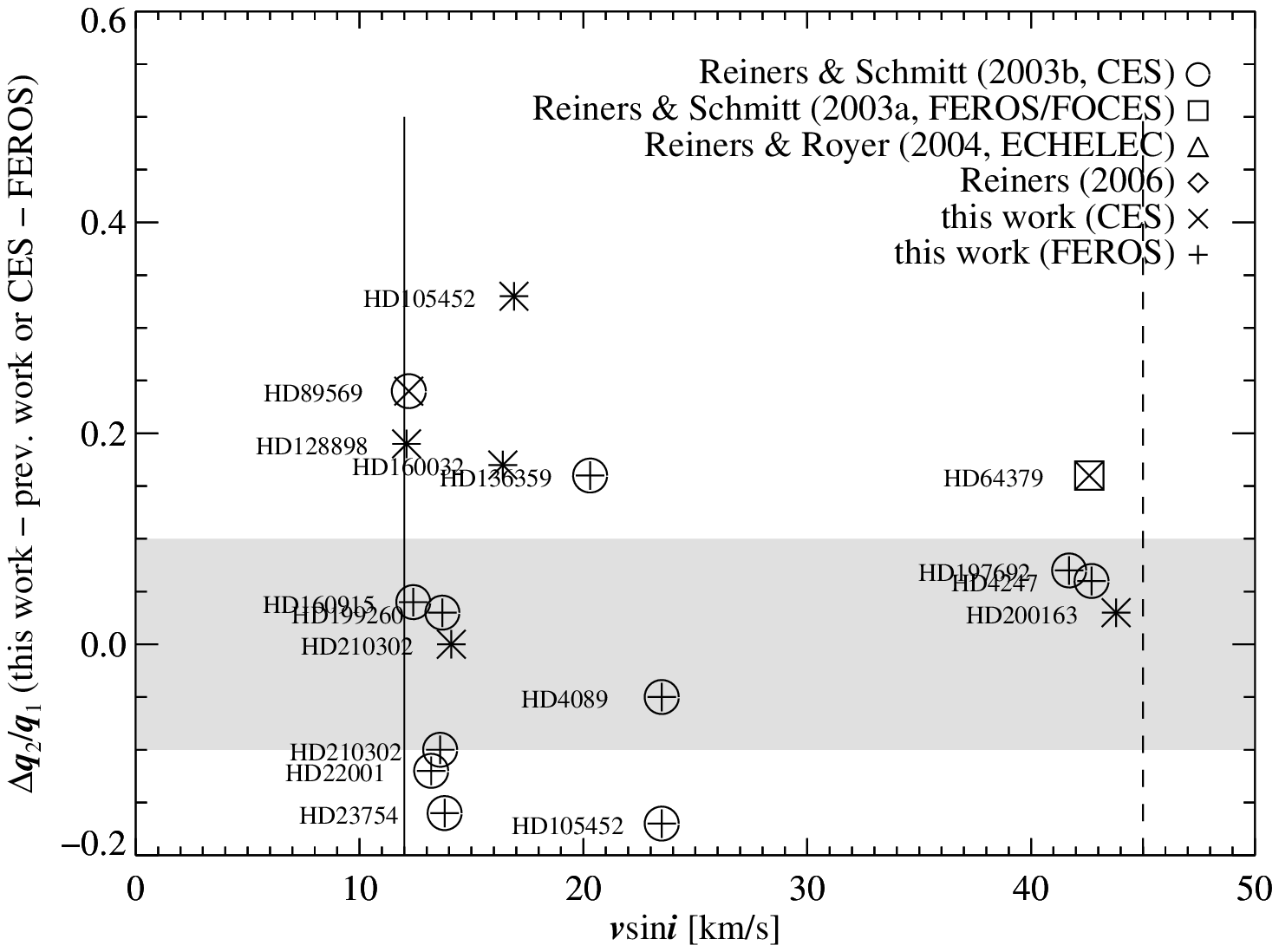}}
\caption{a) $\vsini$ is compared for stars in common to previous and the present work or with both CES and FEROS measurements available (see legend for details). The results mostly agree with differences below $5\,\kmps$. The few outliers are labelled and discussed in the text.  Residuals are calculated following the scheme {\it this work - previous work} and {\it CES - FEROS}. Arrows to the bottom mark upper limits to measurements obtained in the present work or CES measurements of the present work. Arrows to the left denote upper limits to values from previous work or FEROS measurements of the present work. The grey area encompasses all residuals with amounts of less than $5\,\kmps$. The vertical lines indicate the resolution limits of CES (solid) and FEROS (dashed). -- b)  Similar to (a), $\qfrac$ is compared for stars in common with previous studies or with both CES and FEROS measurements available. The grey area harbours all residuals with amounts of less than 0.10. Objects with larger deviations are discussed in the text.} 
\end{figure}
Systematic errors may occur which can not be assessed a priori. This particularly concerns instrumental effects introduced by the spectrographs used. A comparison to the previous results obtained by Reiners et al. for stars in common is helpful as these results are based on data taken at another epoch with different instrumentation. 

In the present work, new CES and FEROS spectra have been taken of stars which have been analysed before by \citet[CES]{\ReinersIIIa} and \citet[FOCES and FEROS]{\ReinersIIIb}. The more recent work of \citet{\ReinersVI} is mostly based on the same spectra. \citet{\ReinersIV} analysed ECHELEC spectra with a resolution even lower than those of FEROS and FOCES.

Figure~\ref{fig:vsini_resid} displays differences between $\vsini$ measurements of the present work and previous work and of measurements from different instruments. While most new measurements agree with previous assessments within a few $\kmps$, discrepancies of more than 5\,$\kmps$ appear in the cases of HD\,80671, HD\,105452, HD\,124425, and \object{HD\,198001} when comparing to \citet{\ReinersIIIb} and \citet{\ReinersIV}, measurements which are based on FEROS, FOCES, and ECHELEC spectra. HD\,80671 and HD\,105452 are particular in that they are spectroscopic binaries. The shape of the rotational profile varies because of the orbital motion of the components. Consequently, the $\vsini$ assessed from different spectra will be different. In particular, the spectra of HD\,105452 analysed in the present work display an asymmetric profile. While \citet{\ReinersIIIb} flagged HD\,80671 as an object with a blended profile, the profile is resolved in the present work but only an upper limit can be determined to the apparently very low intrinsic $\vsini$. HD\,124425 and HD\,198001 were measured at lower spectral resolution in previous work.

Figure~\ref{fig:q2q1_resid_vsini} shows discrepancies of $\qfrac$ measurements together with the corresponding $\vsini$ measurements since these tell us whether a rotational profile can actually be resolved. There are nine residuals larger than 0.10. Only three of them cannot be explained by peculiar line profiles or insufficient spectral resolution. One of these cases is HD\,105452 as can be expected from the comparison of $\vsini$ measurements above. Measurements for \object{HD\,64379} differ by 0.16 which might be related to the fact that the CES spectrum shows a slight asymmetry. The star is a candidate for spottedness according to \citet{\ReinersIIIb}. The profile of \object{HD\,86569} suffers from bad quality in the present work so that the discrepancy of 0.24 with respect to \citet{\ReinersVI} should not be given too much weight and instead the older measurement be preferred. The remaining residuals larger than 0.10 are probably due to the use of different spectrographs. \object{HD\,128898} and \object{HD\,160032} were analysed in the present work based on spectra with different resolution (CES and FEROS) and $\vsini$ is below the sampling limit of FEROS. \object{HD\,22001}, \object{HD\,23754}, and \object{HD\,136359} were analyzed at higher resolving power in previous work.

Generally, in case of profiles without pecularities, discrepancies between different measurements are less than $5\,\kmps$ in $\vsini$ and $0.10$ in $\qfrac$.

\subsection{Distribution of measurements of differential rotation}
\label{sect:distrib}
\begin{figure}
\subfigure[\label{fig:vdistrib}]{\includegraphics[width=\figwidth]{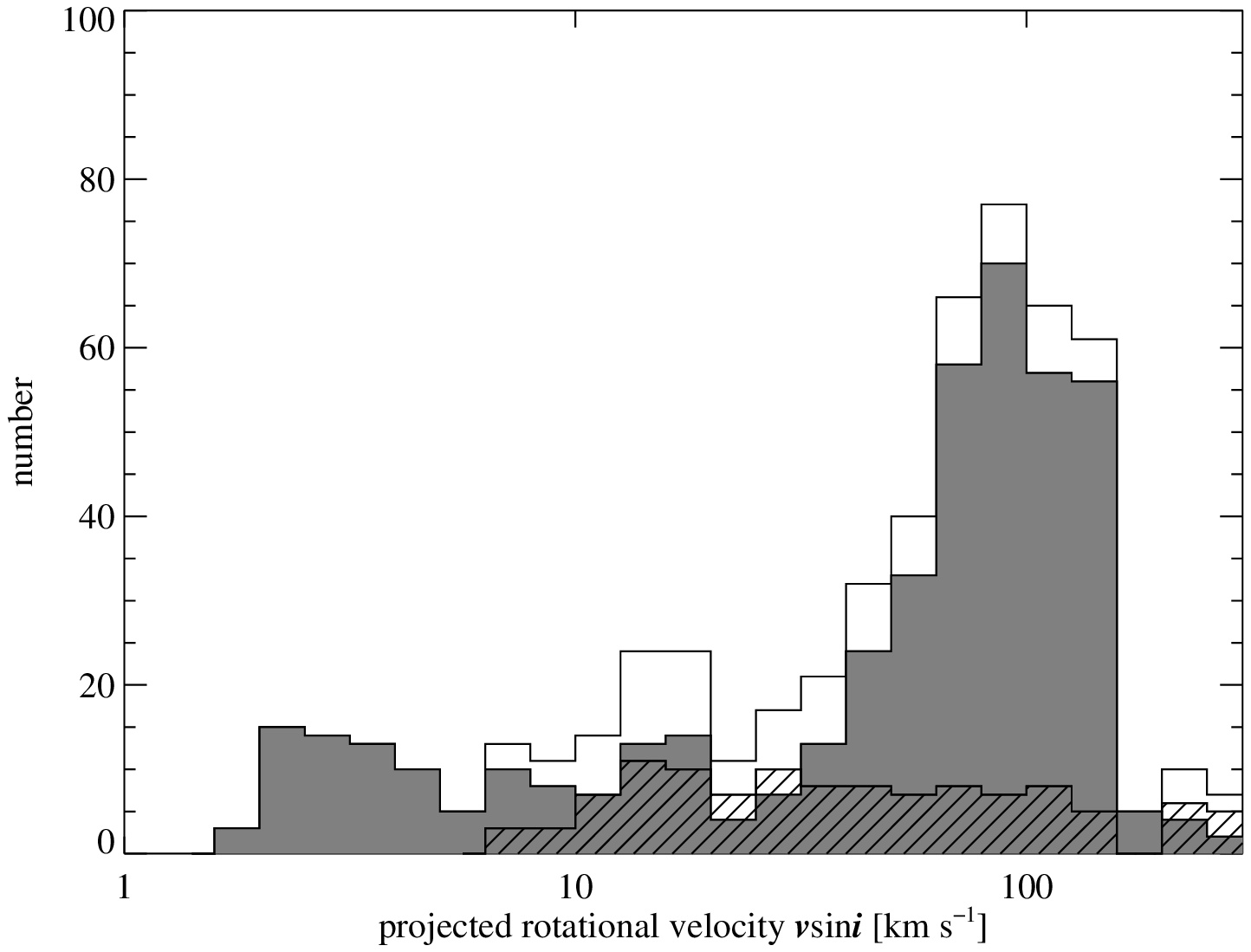}}
\subfigure[\label{fig:qdistrib}]{\includegraphics[width=\figwidth]{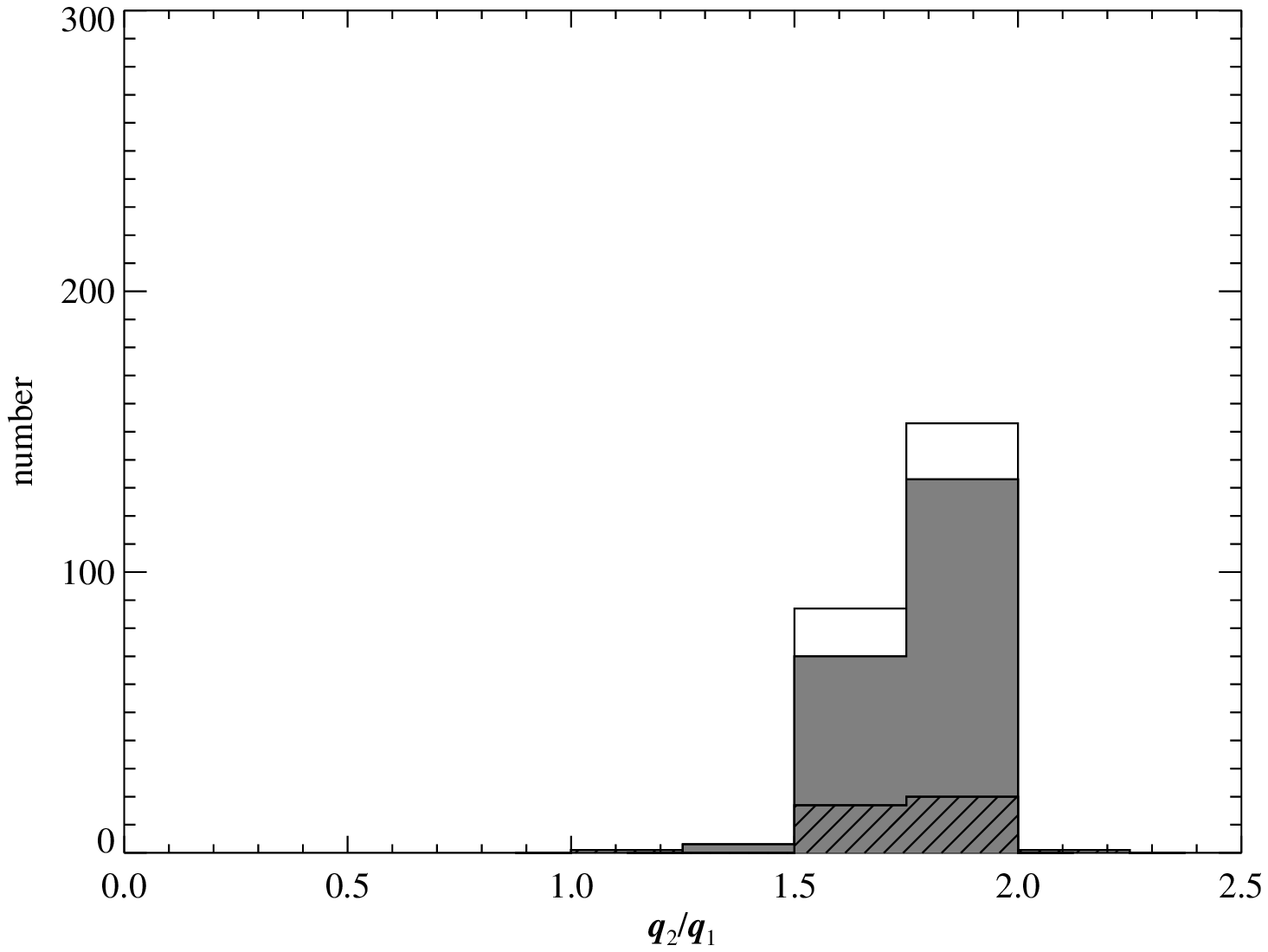}}
\caption{\label{fig:vqdistrib} {\bf (a)} The distribution of projected rotational velocities $\vsini$ is shown for those stars for which measurements could be obtained from the analysis of the zeros of the Fourier transform of the overall broadening profile. Mesurements with upper limits on $\vsini$ are omitted. The grey-shaded area represents the distribution of the sample of the previous work (Reiners {\etal}), the hashed area the sample of the present work. The white area shows the distribution of the combined sample. -- {\bf (b)} The distribution of measurements of $\qfrac$ is shown for the stars with resolved rotational profiles. The layout follows Fig.~\ref{fig:vdistrib}.} 
\end{figure}

Figures~\ref{fig:vdistrib} and \ref{fig:qdistrib} show the distribution of projected rotational velocities $\vsini$ and of the indicator $\qfrac$ of differential rotation for all stars of the sample. In the present work, the rotational profile of 56 stars is studied. In total, there are 300 objects with rotational profile studied by line profile analysis, also yielding precise projected rotational velocities $\vsini$. Many of the stars studied in the present work display line profiles which cannot be studied by line profile analysis because of asymmetries caused by spots or multiplicity. However, information on $\vsini$ can still be obtained. $\vsini$ is inferred for 68 stars and upper limits for further 68 objects. These stars are no longer considered in the present work but the $\vsini$ measurements are tabulated in Table~\ref{tab:results}.

In the present work, ten stars show clear signatures of differential rotation between $\alpha=10$\,\% and 54\,\%. While three of these are new discoveries, seven objects belong to the samples of \citet{\ReinersIIIa} and \citet{\ReinersIIIb} and are now identified or confirmed based on high-resolution CES spectra. Including previous work, there are 33 differential rotators detected by line profile analysis (excluding those possibly affected by multiplicity). 

Three further stars with extremely fast rotation show line profiles indicative of differential rotation. However, in these cases, the feature is plausibly caused by gravitational darkening in the regime of rigid rotation (see Sect. \ref{sect:alpha}).

The amount of differential rotation of six stars is below 6\,\% which is below our detection limit and consistent with rigid rotation. Yet, as is the case for many other objects with undetected differential rotation, small rates of (relative) differential rotation -- and even strong amounts of horizontal shear as is discussed later on -- are still possible. Two F stars (\object{HD\,104731} and HD\,124425) with $T_\mathrm{eff}\approx6500\,$K and $\vsini\lesssim30\,$km\,s$^\mathrm{-1}$ display the lowest values of $\frac{q_\mathrm{2}}{q_\mathrm{1}}$ measured so far by line profile analysis corresponding to differential rotation rates of 54\,\% and 44\,\%, respectively. Five stars show signatures which can be explained by anti-solar differential rotation or more plausibly by rigid rotation with cool polar caps. In total, there are 43 stars with spectra consistent with rigid rotation.

\section{Rotation and differential rotation in the HR diagram}
\label{sect:HRD}
The distribution of differential rotators in the HR diagram is of particular interest since it relates the frequency of differential rotation to their mass and evolutionary status. 

To construct the HR diagram, homogeneous photometric and stellar data are used for the sample of the present work. $(B-V)$ colours, $V$ band magnitudes and parallaxes were drawn from the Hipparcos catalogue when available and from Tycho-2 otherwise. Bolometric corrections and effective temperatures $\Teff$ were calculated from $(B-V)$ using the calibration of \citet{\Flower}. Absolute $V$ band magnitudes $M_\mathrm{V}$ were inferred from the apparent magnitudes and the parallaxes. Radii are calculated from effective temperature and bolometric magnitude which is derived from absolute $V$ band magnitude. 

The study of differential rotation throughout the HR diagram done in the present work includes previous analyses by Reiners {\etal} (2003-2006). Not all photometric and basic stellar data used in the present work are equally available from there. Therefore, data needs to be complemented as consistently as possible and without changing results previously obtained. Therefore, photometry was adopted from Reiners {\etal} (2003-2006) when available and otherwise completed according to the present work. Stellar parameters of previously analyzed stars are adopted from Reiners {\etal} when available. Otherwise, they are derived in the same way as described above for the data of the present sample. Table~\ref{tab:comp_diffrot} presents rotational data and basic stellar data for all stars studied by line profile analysis. 

Sometimes the stars have been studied more than once by line profile analysis. It is not always the most recent result that has been adopted in Table~\ref{tab:comp_diffrot}. If possible, the most plausible and conclusive result is used, for example in cases of peculiar profiles when the line profile could be resolved in separate components later on. Generally, data are preferentially adopted which are based on spectra with higher resolution. Also, data with more supplemental information given by flags are preferred.

\begin{figure}
\subfigure{\includegraphics[width=\figwidth]{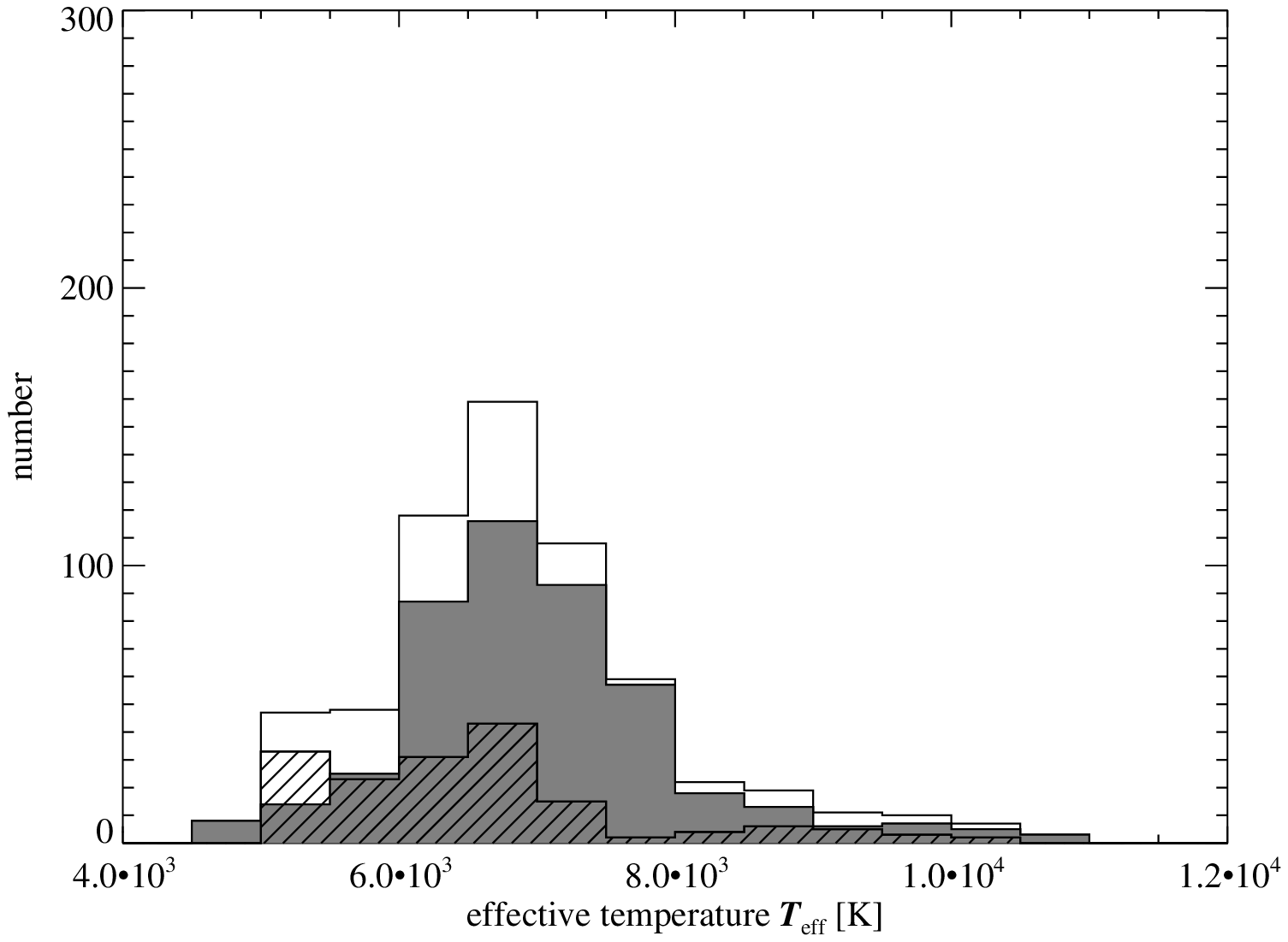}}
\subfigure{\includegraphics[width=\figwidth]{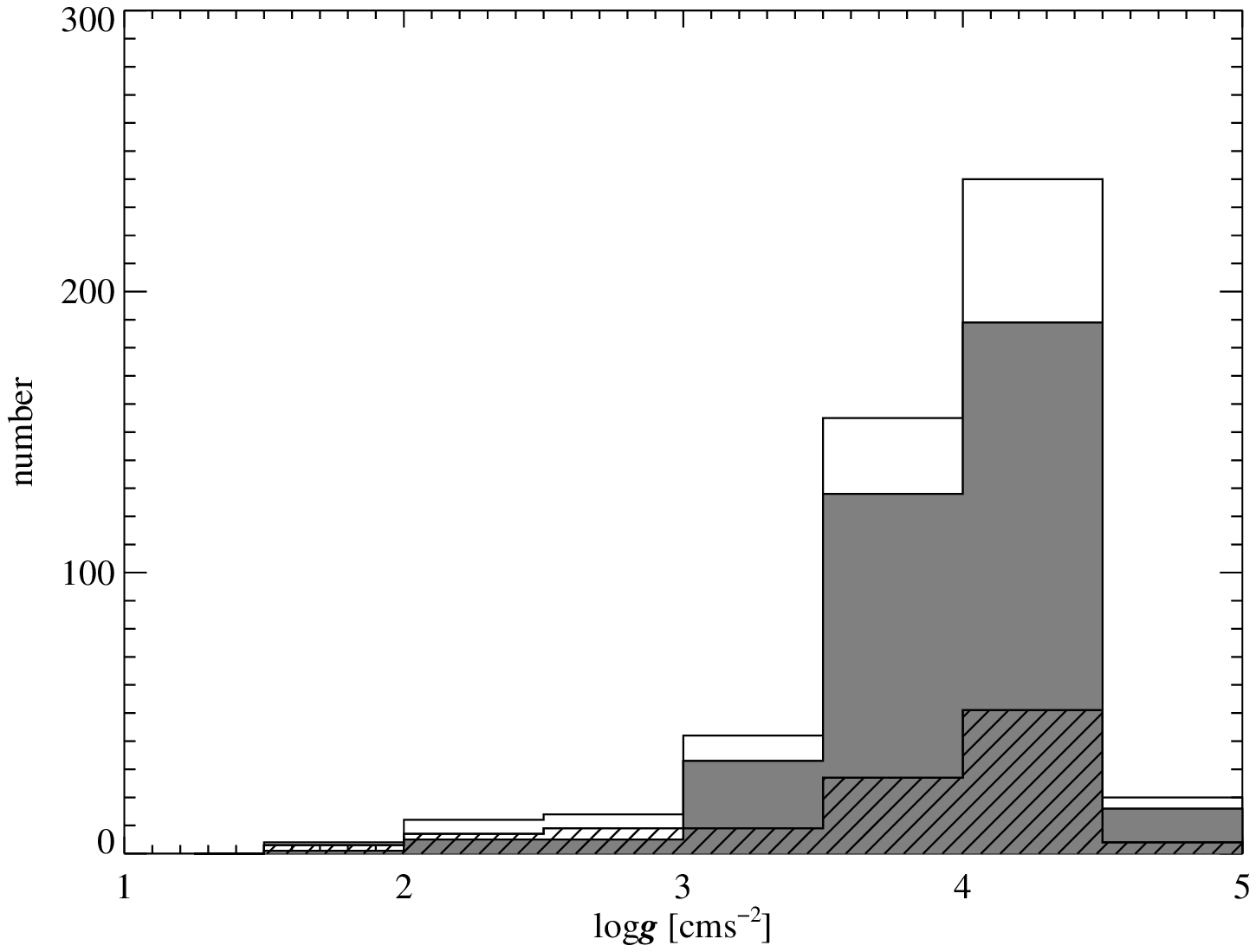}}
\caption{\label{fig:tgdistrib} The distributions of effective temperature and surface gravity are shown. The layout follows Fig.~\ref{fig:vqdistrib}.} 
\end{figure}

Approximate surface gravities for all stars are inferred from photometry and from mass estimates which in turn are derived from effective temperatures using the calibration for dwarf stars of \citet[][appendix B]{Gray05}. Thus, the surface gravities will be upper limits only for giants and lower limits in the case of unresolved binaries. The distribution of effective temperature and surface gravity is shown in Fig.~\ref{fig:tgdistrib}.

\begin{figure}
\includegraphics[width=\figwidth]{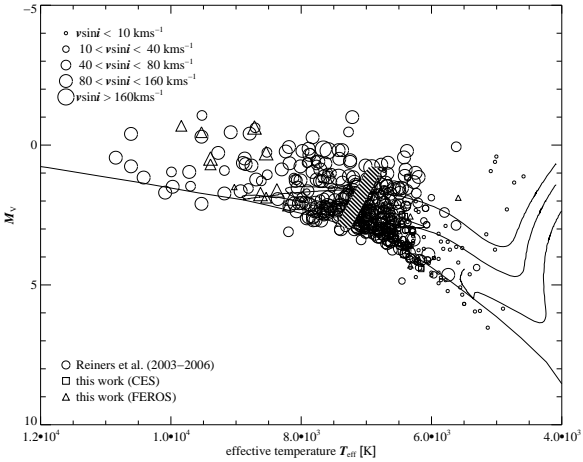}
\caption{\label{fig:hrd_vsini}HR diagram with the stars from previous (circles) and the present (squares and triangles) work. Symbol size scales with projected rotational velocity measured by the FT method. The granulation boundary according to \citet{\Gray} is indicated by the hashed region. Evolutionary tracks for 1.0, 1.5, and 2.0\,$\Msun$ and the early-main sequence of \citet[][solid, $Z=0.02$ without overshooting]{\Siess} are added, using bolometric corrections by \citet[][table~A5]{\Kenyon}. Stars with photometry possibly affected by multiplicity (flags 'x' and 'm' in Tables~\ref{tab:results} and \ref{tab:comp_diffrot}) are not included.} 
\end{figure}
Not all parts of the HR diagram are equally accessible to line profile analysis. Figure~\ref{fig:hrd_vsini} shows the distribution of the analyzed stars in the HR diagram in terms of rotational speed. The location of the stars is compared to evolutionary models of \citet{\Siess} and the granulation boundary according to \citet{\Gray}. The granulation boundary indicates the location where deep convective envelopes form and the approximate onset of magnetic braking. Accordingly, the figure illustrates that Sun-like and cooler main-sequence stars are efficiently braked. Also the late-type giant stars display slow rotation. Consequently, these and most cool dwarf stars are not accessible to the study of the rotational profile. Therefore, the present study is generally restricted to main-sequence and slightly evolved stars of spectral types A and F.

\subsection{Two populations of differential rotators}
\label{sect:populations}
\begin{figure*}
\includegraphics[width=\textwidth]{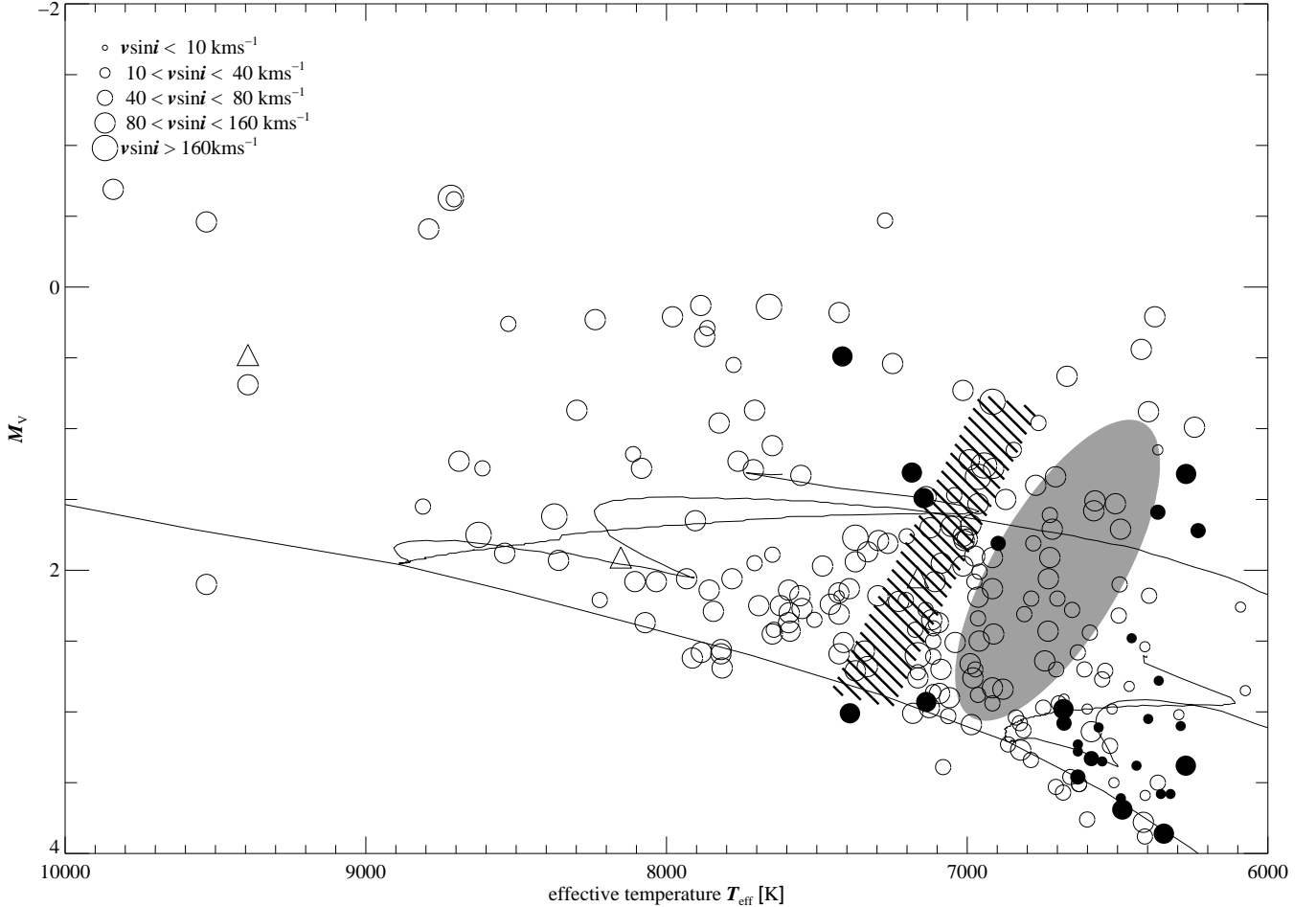}
\caption{\label{fig:hrd_diffrot} The HR diagram with the data of the present work and the results from Reiners {\etal} (2003-2006). Differential rotators are indicated by filled circles while rigid rotators are shown by open circles. Evolutionary models and the granulation boundary are indicated as in Fig.~\ref{fig:hrd_vsini}. The shade highlights the region where no differential rotators are detected. Triangles denote \object{HD\,30739}, \object{HD\,43940}, \object{HD\,129422} which are considered rigid rotators with gravitational darkening (Sect.~\ref{sect:HRD}). The fast differential rotators at the granulation boundary are \object{HD\,6869}, \object{HD\,44892}, \object{HD\,60555}, \object{HD\,109238}, and \object{Cl*IC\,4665\,V\,102} which display strongest horizontal shear \citep{\ReinersIV,\ReinersVI}. Stars possibly affected by multiplicity (flags 'm', 'x', and 'y' in Tables~\ref{tab:results} and \ref{tab:comp_diffrot}) are not included.}
\end{figure*}

The HR diagram in Fig.~\ref{fig:hrd_diffrot} shows that the present work adds several rigid rotators in the hot star region between 8000 and 10000\,K. Three stars are detected -- the A stars HD\,30739, HD\,43940, and HD\,129422 at the transition to spectral type F -- with line shapes indicative of differential rotation. However, the low values of $\qfrac$ might be caused to a certain extent by gravitational darkening in the regime of rigid body rotation since all three objects rotate very rapidly with $\vsini>200\,\kmps$. This scenario is suggested by \citet{\ReinersIV} for the rapid rotator HD\,44892 which is the hottest and most luminous differential rotator in Fig.~\ref{fig:hrd_diffrot}. In the cases of HD\,6869, HD\,60555, and HD\,109238 they argue, however, that this cannot be the sole explanation for the low values of $\qfrac$ since rotational speed would exceed breakup velocity. These stars are all located at the granulation boundary and display the strongest absolute horizontal shear \citep{\ReinersIV}. There are two more differential rotators close to the granulation boundary at its cool side, Cl*\,IC\,4665\,V\,102 and \object{HD\,72943} \citep{\ReinersVI}. V\,102 displays the strongest horizontal shear $\Delta\Omega$ of the whole sample. \citet{\ReinersVI} argue that a mechanism may be responsible for the strong shear at the granulation boundary which is different from the mechanism at work in stars with deeper convective envelopes at cooler effective temperatures.

At early-F spectral types, there is evidence for a lack of differential rotators (shaded area in Fig.~\ref{fig:hrd_diffrot}). At later F and early-G spectral types, however, a dense population of differential rotators exists on the main sequence and a scattered population towards higher luminosity. At these later spectral types, there are HD\,104731 and HD\,124425, the stars with the highest values of relative differential rotation $\alpha$ measured by line profile analysis. The object with strongest horizontal shear in this region, however, is HD\,64185\,A and  it is the only object there that displays a shear strength comparable to the stars at the granulation boundary. This result has to be regarded with care since HD\,64185\,A is a component of a spectroscopic binary, or possibly a triple (Sect.~\ref{sect:mult}).

\subsection{Dependence of relative differential rotation on stellar parameters}
\label{sect:depend}

\begin{figure}
\subfigure[\label{fig:compt}]{\includegraphics[width=\figwidth]{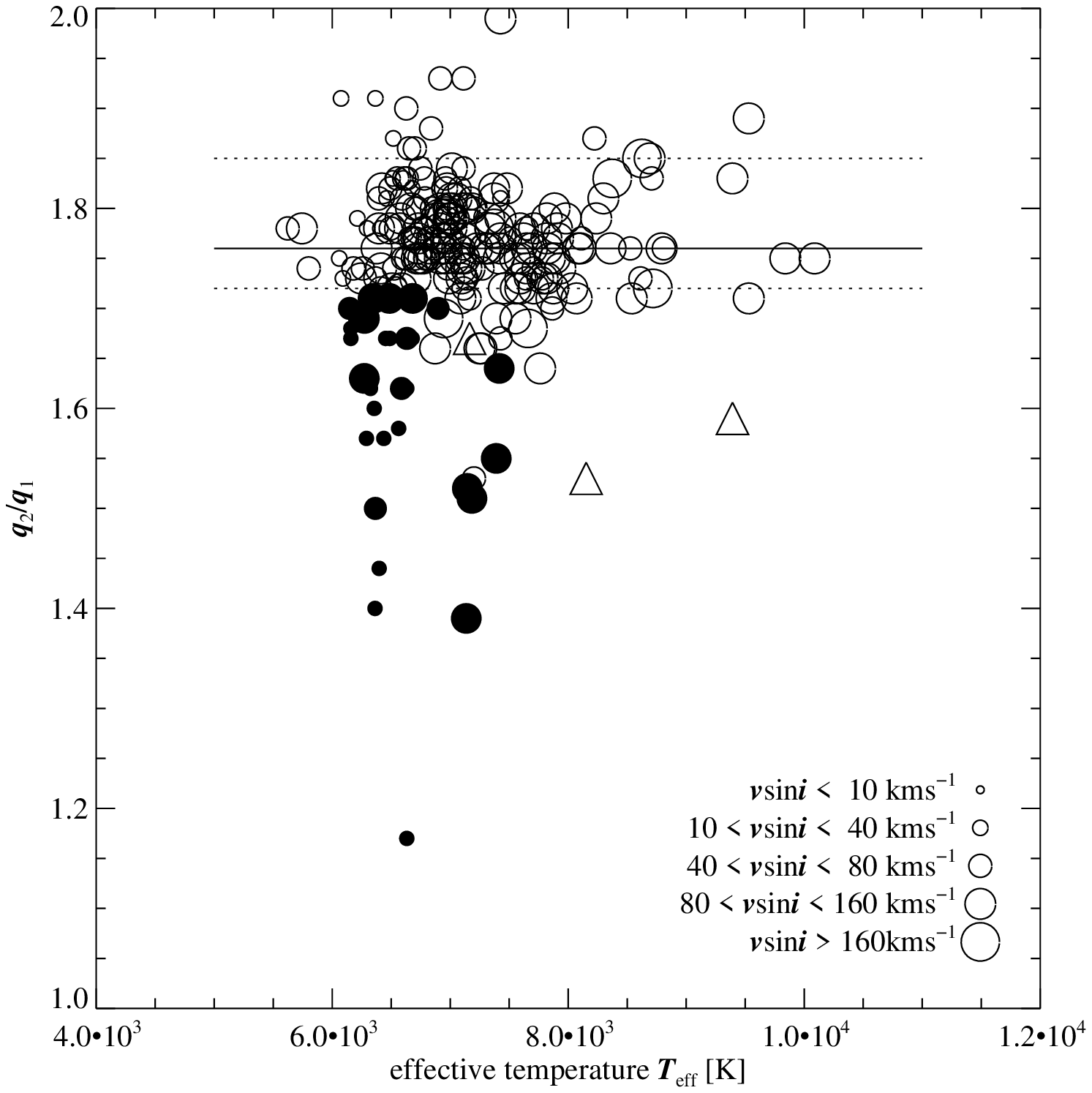}}
\subfigure[\label{fig:tfrac}]{\includegraphics[width=\figwidth]{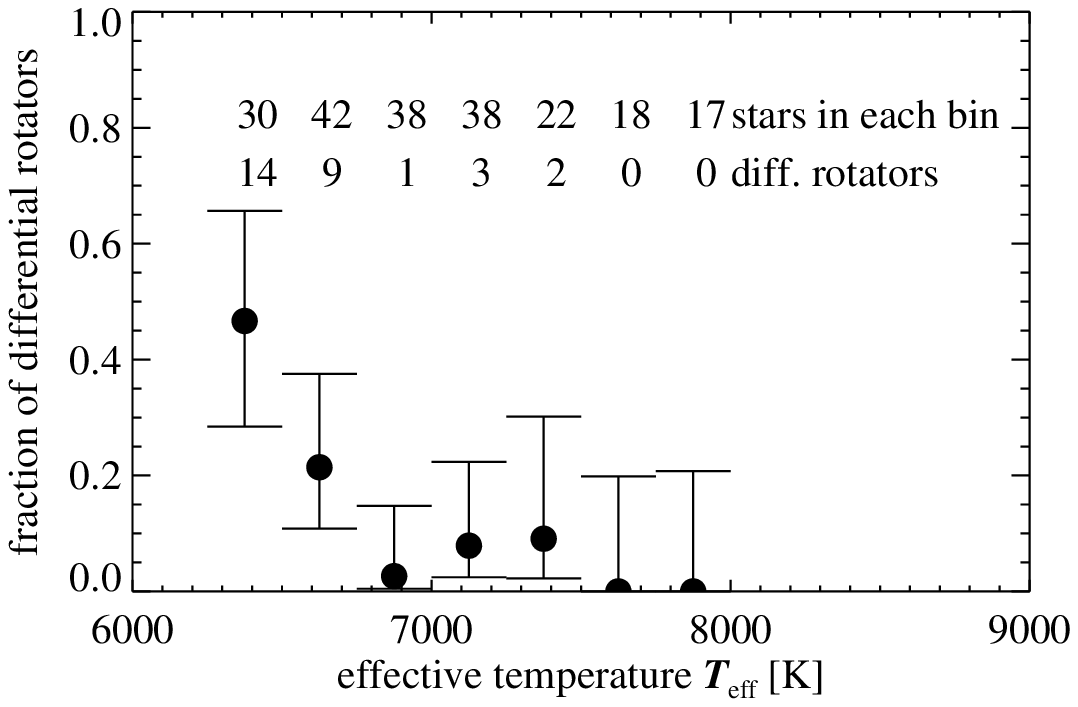}}
\caption{a) The ratio $\qfrac$ indicates the amount of differential rotation and is plotted here vs. effective temperature $\Teff$. Dotted lines indicate the range of $\qfrac$ for rigid body rotation and unknown limb darkening. The solid line denotes $\qfrac$ for the case of rigid rotation and Sun-like limb darkening ($\varepsilon=0.6$) Symbol size scales with projected rotational velocity $\vsini$ as in Fig.~\ref{fig:hrd_vsini}. Open circles denote rigid rotators and filled circles differential rotators. The triangles show the cases where the low fraction of $\qfrac$ is probably caused by gravitational darkening. The data point with lowest $\qfrac$ represents HD\,104731 which displays the strongest relative differential rotation discovered in the present work. The lack of differential rotators at early-F types can be spotted easily. -- b) $2\sigma$ confidence limit on the fraction of differential rotation with respect to effective temperature. Only bins with a number of at least 15 objects are displayed.} 
\end{figure}

Fig.~\ref{fig:compt} displays the basic quantity measured, i.e. the ratio of the zeros of the Fourier transform $\qfrac$, vs. effective temperature $\Teff$, including measurements of Reiners {\etal} (2003-2006). Values of $\qfrac$ between 1.72 and 1.85 are consistent with rigid body rotation when accounting for unknown limb darkening. The value of 1.76 represents rigid body rotation when assuming a linear limb darkening law with a Sun-like limb darkening parameter $\varepsilon=0.6$. In the regime of rigid rotation and marginal solar/anti-solar differential rotation, the present work does not change the overall distribution found by Reiners {\etal} but the lack of early-F type differential rotators between 6700 and 7000\,K becomes more pronounced. Nevertheless, the strongest differential rotators are located close to this gap. 

The figure corroborates evidence of two different populations of differential rotators, one with moderate to rapid rotation at the cool side of this gap and one with extremely rapid rotation at the granulation boundary at the hot side. At the cool side, stars are generally rotating more slowly than stars on the hot side of the gap.

On average, the fraction of differential rotators among stars with known rotational profiles significantly decreases with increasing effective temperature (Fig.~\ref{fig:tfrac}). Those fractions are estimated from the number counts in bins of effective temperature. It is assumed that the number counts originate from a binomial probability distribution with an underlying parameter $p$ which is the probability that a randomly chosen star among the stars with measured rotational broadening is a differential rotator. From this distribution and the actual number counts of differential rotators in each temperature bin, $2\sigma$ confidence intervals on the true value of $p$ are derived \citep[following][]{Hengst67}.

\begin{figure}
\subfigure[\label{fig:compv}]{\includegraphics[width=\figwidth]{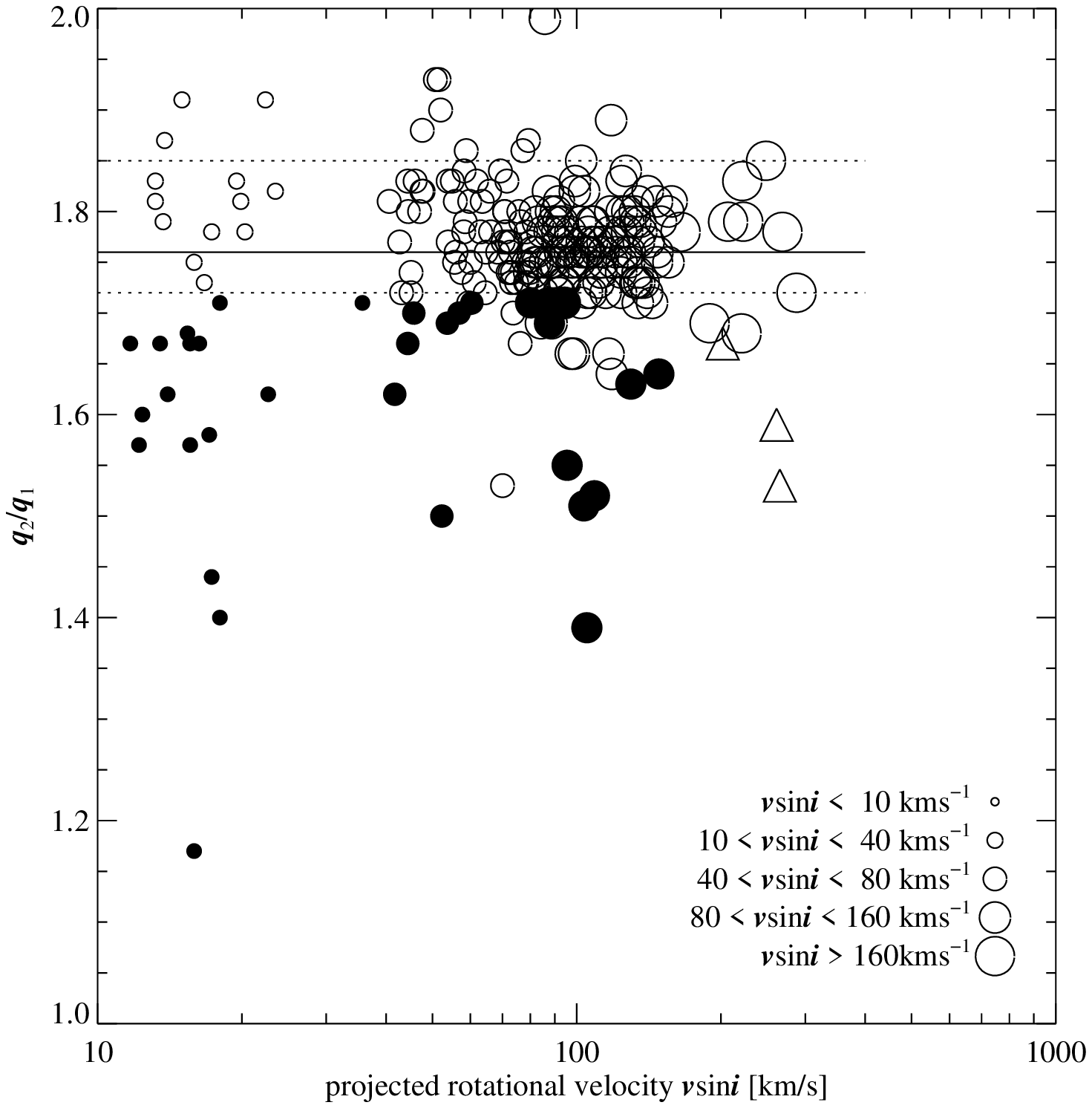}}
\subfigure[\label{fig:vfrac}]{\includegraphics[width=\figwidth]{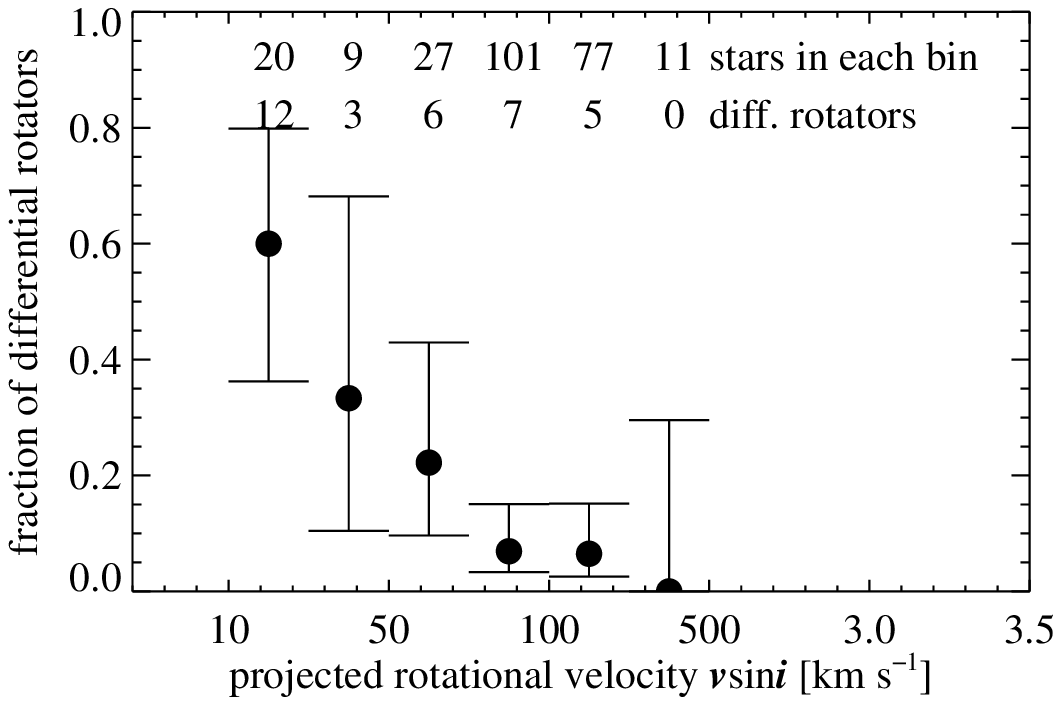}}
\caption{a) In a way similar to Fig.~\ref{fig:compt}, $q_2/q_1$ is plotted vs. the projected rotational velocity $\vsini$. Symbols are the same as in Fig.~\ref{fig:compt}. Although $\vsini$ are given by the bottom axis the $\vsini$ scaling of symbol size is kept to facilitate the identification of data points when comparing to Fig.~\ref{fig:compt}. -- b) $2\sigma$ confidence limit on the fraction of differential rotation with respect to projected rotational velocity.} 
\end{figure}

Fig.~\ref{fig:compv} displays the ratio of the zeros of the Fourier transform $\qfrac$ vs. the projected rotational velocity $\vsini$ in a way similar to Fig.~\ref{fig:compt}. The dependence of the frequency of differential rotation on $\Teff$ and $\vsini$ cannot be studied independently since $\vsini$ and $\Teff$ are strongly correlated and thus degenerate in what concerns their effect on differential rotation \citep{\ReinersVI}. Therefore, it is not a surprise, that Fig.~\ref{fig:compv} looks very similar to Fig.~\ref{fig:compt}. Differential rotation is common among slow rotators and becomes rare at rapid rotation (see Fig.~\ref{fig:vfrac}). From Fig.~\ref{fig:compt} and \ref{fig:compv} one notices that almost all differential rotators at the granulation boundary are fast rotators with $\vsini\gtrsim100\,\kmps$.

\begin{table*}[!ht]
\caption{\label{tab:frac}Fraction of hot, giant, and rapidly rotating stars among differential and rigid rotators. These fractions are compared to the fractions within the combined sample.}
\begin{center}
{\small 
\begin{tabular}{lccc}
\noalign{\smallskip}
\hline\hline
\noalign{\smallskip}
&differential rotators&rigid rotators&combined\\
\noalign{\smallskip}
\hline
\noalign{\smallskip}
fraction of hot stars ($T_\mathrm{eff}\,>\,7000\,$K)&	$45^{+18}_{-17}$\,\%&$88^{+4}_{-6}$\,\%&$82^{+5}_{-6}$\,\%\\
\noalign{\smallskip}
fraction of giant stars ($\log{g}\,<\,3.5$)&$12^{+17}_{-8}$\,\%&$11^{+6}_{-4}$\,\%&$12^{+5}_{-4}$\,\%\\
\noalign{\smallskip}
fraction of rapid rotators ($\vsini\,>\,50$\,km\,s$^{-1}$)&$42^{+19}_{-17}$\,\%&$88^{+4}_{-6}$\,\%&$81^{+5}_{-6}$\,\%\\
\noalign{\smallskip}
\hline
\end{tabular}
}
\end{center}
\end{table*}

The strength of differential rotation, in contrast to the frequency of differential rotation, does not vary for different values of effective temperature or projected rotational velocity.  The lowest values of $\qfrac$ measured tend to be lower at cool effective temperature and slow rotation but these trends are not significant. 

In summary, the fraction of differential rotators decreases significantly with increasing $\Teff$ and $\vsini$. In other words, the fraction of hot and rapidly rotating stars among differential rotators is significantly less than among rigid rotators (see Table~\ref{tab:frac}).

\begin{figure}
\subfigure[\label{fig:compg}]{\includegraphics[width=\figwidth]{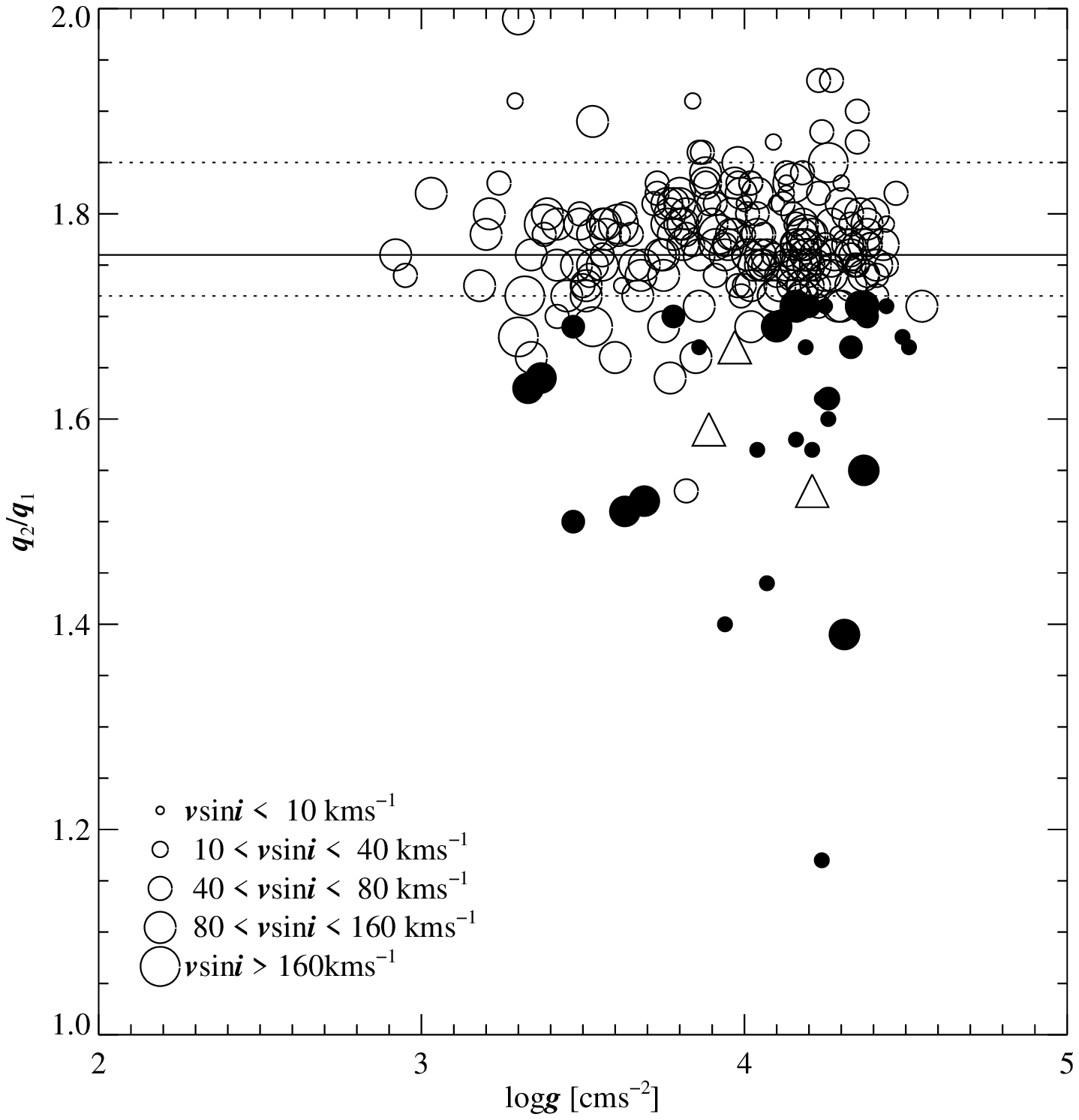}}
\subfigure[\label{fig:gfrac}]{\includegraphics[width=\figwidth]{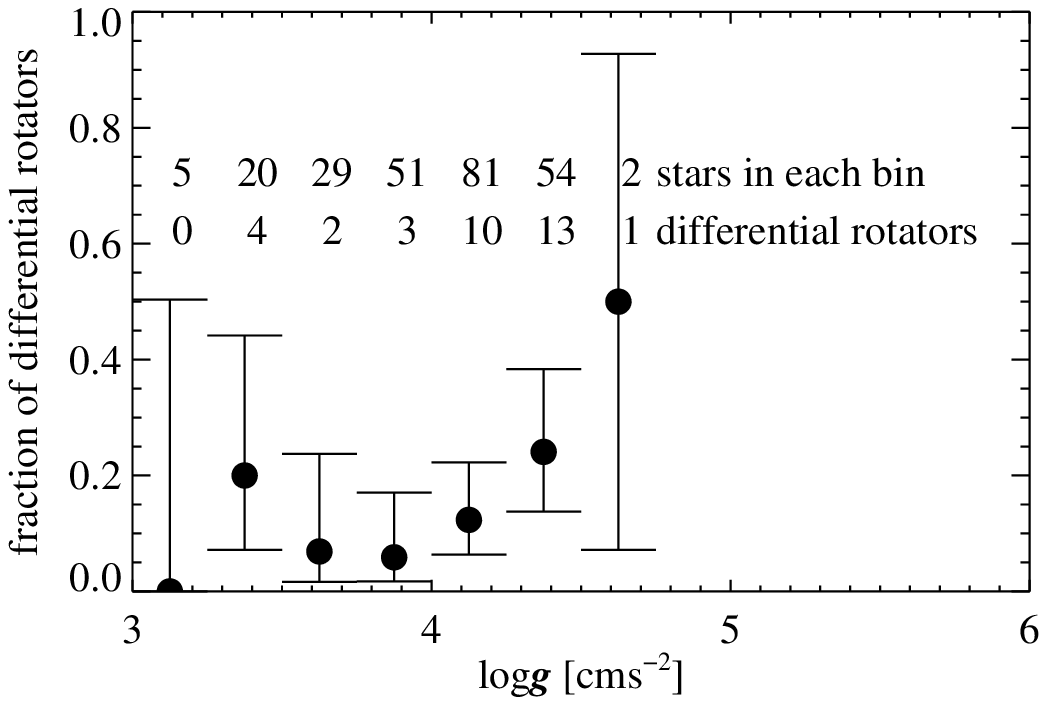}}
\caption{a) In a way similar to Fig.~\ref{fig:compt}, $q_2/q_1$ is plotted vs. the surface gravity $\log{g}$. Symbols are the same as in Fig.~\ref{fig:compt}. -- b) $2\sigma$ confidence limit on the fraction of differential rotation with respect to surface gravity.} 
\end{figure}

In terms of surface gravity, the fraction of differential rotators increases towards high gravity but with the data at hand, differences are not significant (Figs.~\ref{fig:compg}, \ref{fig:gfrac}). There are roughly as many giants among the differential rotators as among the rigid rotators and the whole sample. We finally recall the reader that the sample is comprehensive but incomplete and biased towards the selection of suitable targets.

\section{Discussion of horizontal shear}
\label{sect:shear}

\subsection{Measured shear and rotational period}
\label{sect:shear_period}
\begin{figure}
\includegraphics[width=\figwidth]{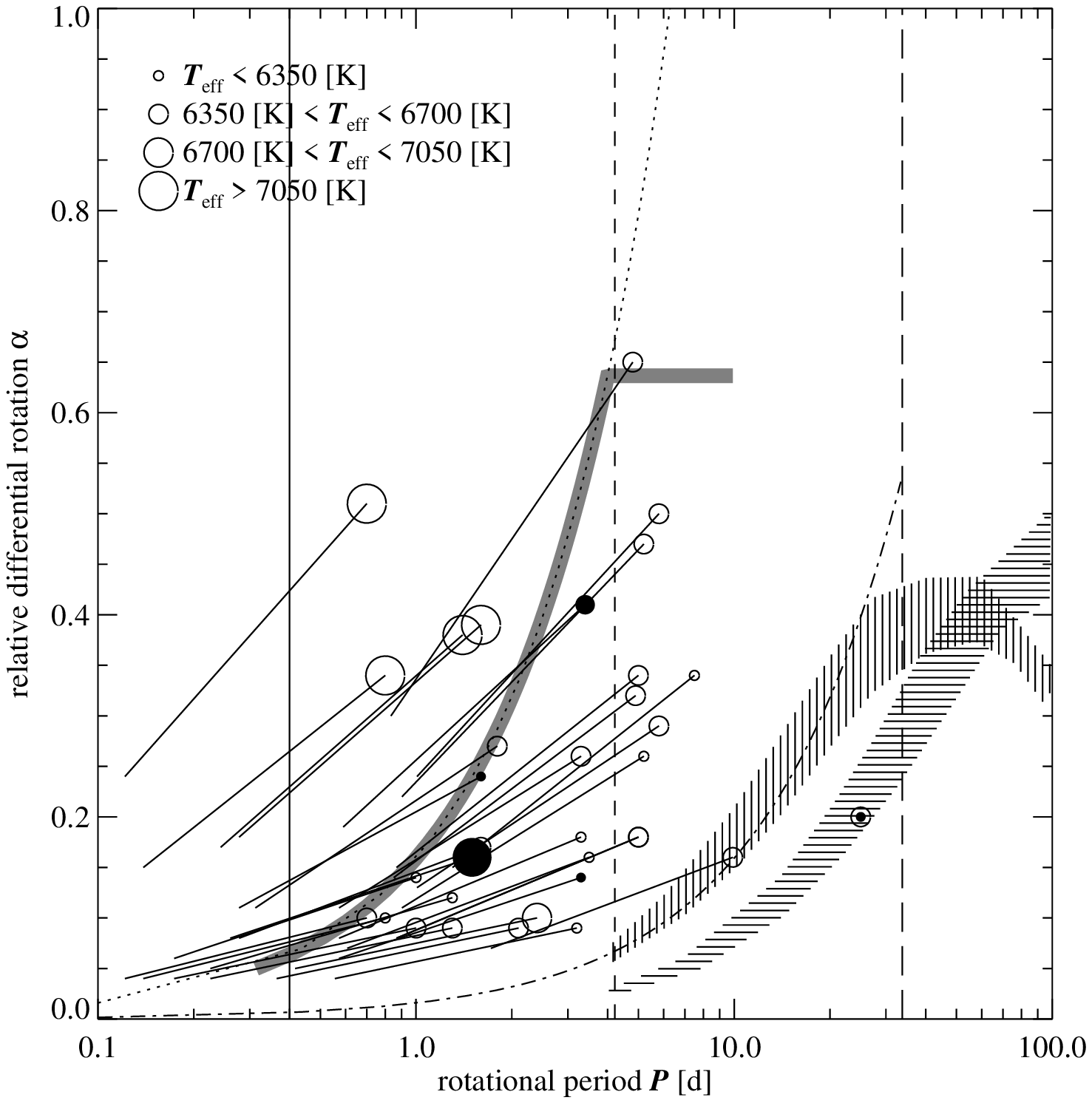}
\caption{\label{fig:alpha_P}The figure displays relative differential rotation $\alphaimax$ vs. estimated period for all differential rotators detected by profile analysis. The dotted and dashed-dotted lines indicate the dependence of $\alpha$ on period for constant absolute shear $\Delta\Omega=1.0\,\radpd$ and $0.1\,\radpd$, resp. Open symbols indicate dwarfs and filled symbols evolved stars with ${\log}g\leq3.5$. Symbol size scales with effective temperature as indicated in the legend. The straight solid lines relate the data points to the values derived for 10$\deg$ thus illustrating the error due to inclination. Errors intrinsic to $\alphaimax$ and $\alphaimin$ are omitted for clarity but might be substantial. The scaling of symbol size with effective temperature allows one to compare with the theoretical predictions for an F8 star (vertically hashed) and a G2 star (horizontally hashed) \citep[][width of relations given by cases $c_\mathrm{v}=0.15$ and 4/15]{\KuekerV}. The vertical lines indicate period limits accessible to measurement. The vertical solid line gives the shortest rotational period of the stars with a measurement of the rotational profile while the other vertical lines denote the periods corresponding to the $\vsini$ limit of CES for a dwarf star (short-dashed) of 1 solar radius and an evolved star (long-dashed) of 8 solar radii. The FEROS limit will be at even shorter periods. The location of the Sun is also shown ($\odot$). The grey solid line gives the upper envelope to mid and late-F type differential rotators discussed in the text.} 
\end{figure}

\begin{figure}
\includegraphics[width=\figwidth]{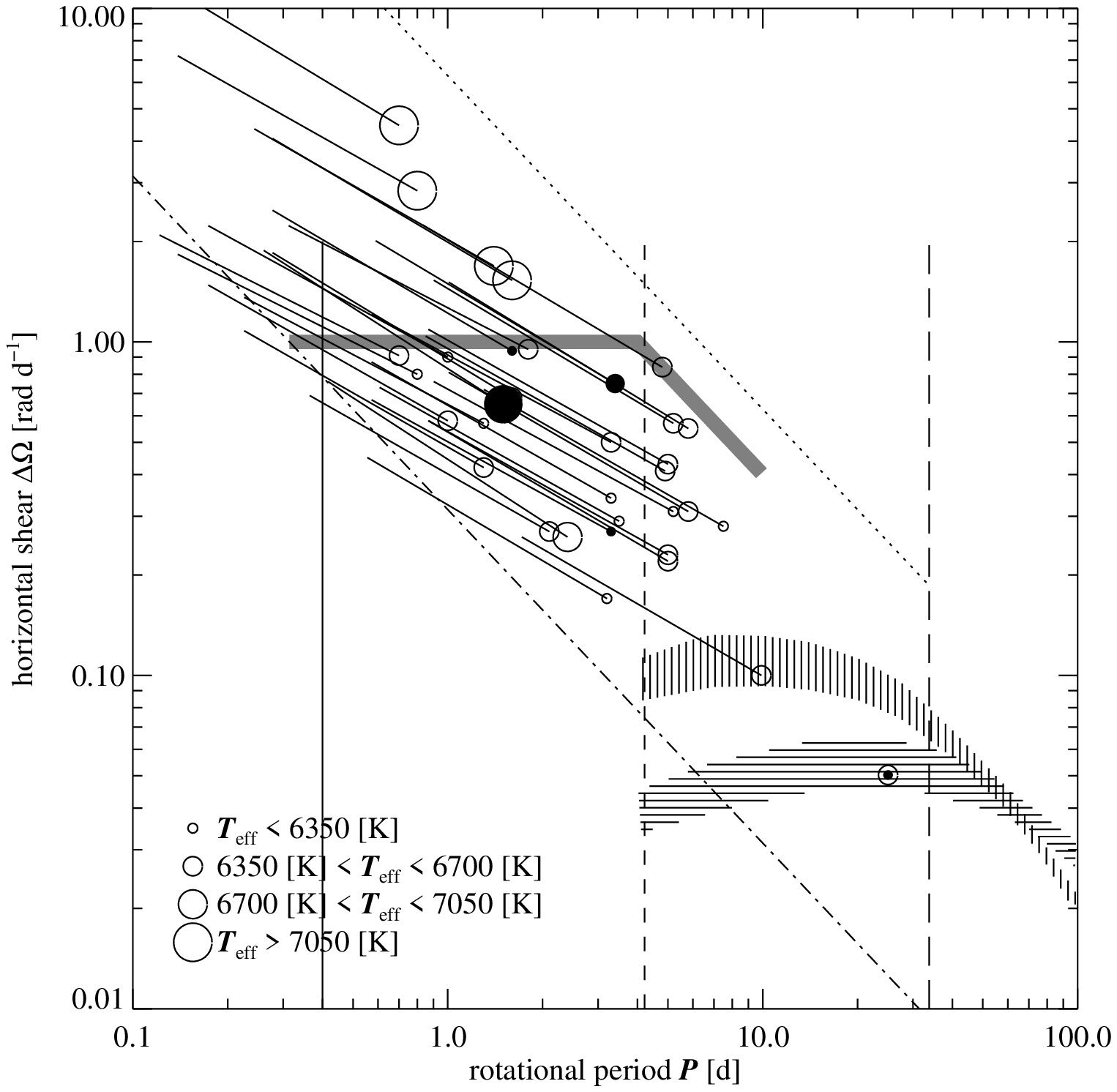}
\caption{\label{fig:shear_P}The figure displays the absolute shear $\Delta\Omegaimax$ vs. estimated projected rotation period. The dash-dotted and dotted lines indicate absolute shear for constant $\alpha=0.05$ (detection limit) and 1.0, resp. The other curves and the symbols have have the same meaning as in Fig~\ref{fig:alpha_P}.} 
\end{figure}

In order to study the strength of differential rotation, the measured quantity $\qfrac$ is converted to relative differential rotation $\alpha$ as described in Sect.~\ref{sect:alpha} and via rotational period to absolute horizontal shear $\Delta\Omega$ (Eq.~\ref{eq:shear}). Rotation periods are given in Table~\ref{tab:comp_diffrot} and are estimated from the $\vsini$ derived and the photometric radius.

When interpreting the measured relative differential rotation $\alpha$, unknown inclination certainly is one of the most important uncertainties.
Unknown inclination effectively turns the measurements of $\alpha$ into upper limits because $\qfrac$ is an approximate function of $\alphasini$.
However, assuming a uniform distribution of inclination angles, and according to the statistics of a $\sini$ distribution, much less stars will have low values of $\sini$ while most will have $\sini\lesssim1$.
Although the uncertainties introduced by unknown limb darkening and the measurement error of $\qfrac$ might be substantial, only the error bar due to inclination is indicated here (represented by the values $\alphaimax$ and $\alphaimin$ introduced in Sect.~\ref{sect:alpha}).

Figs.~\ref{fig:alpha_P} and \ref{fig:shear_P} present the relation of differential rotation with rotational period. The measurements from line profile analysis cover rather short periods of $\approx1-10$\,d. This is not surprising since the study only accesses rapid rotators. Relative differential rotation ranges from the detection limit at 5\,\% to values of more than 60\,\% when assuming an incliation angle of 90\,$\deg$. This corresponds to absolute horizontal shear of the order of $0.1-1\,\radpd$.

At each rotational period, a different range of horizontal shear is accessible as given by the parallel dotted and dash-dotted lines in Fig.~\ref{fig:shear_P}. The detection limit of $\alpha$ corresponds to a minimum detectable shear which decreases with increasing period, e.g. a shear of $\approx0.01\,\radpd$ can be detected at a period of about 30\,d and above. The line indicating $\alpha=100\,$\% corresponds to the extreme case that the difference of angular velocity between the poles and the equator is as large as the equatorial rate of rotation. If this is considered as maximum attainable relative differential rotation, no shear higher than $\approx0.2\,\radpd$ will be seen at periods larger than about 30\,d, for example. Substantial amounts of shear of the order of $1\,\radpd$ and more can only be seen at periods shorter than $\approx10\,$d. It is worth to note that line profile analysis allows one to detect weakest horizontal shear at long rotation periods.
Then, the measurement is only limited by $\vsini$. 

The accessible range in horizontal shear is not fully covered with measurements. Certainly, at long periods, the corresponding $\vsini$ will be too slow for the measurement by line profile analysis as is indicated by the vertical lines in Fig.~\ref{fig:alpha_P} and \ref{fig:shear_P} for a dwarf and a giant star. At short periods of the order of 1\,d, some structure becomes apparent which is related to the frequency distribution in the HR diagram discussed above.
In detail, two different groups of differential rotators stand out. The first group comprises the rapid differential rotators at the granulation boundary. The second group consists of the cooler differential rotators. Both groups cover values of relative differential rotation of similar magnitude (Figure~\ref{fig:alpha_P}) but the second group is located at longer periods. Figure~\ref{fig:shear_P}, however, shows that the first group of stars displays stronger shear than the second group. \citet{\ReinersVI} already noticed that the objects at the granulation boundary have high $\vsini$ around $100\,\kmps$ and that all of these cluster at short periods at high horizontal shear. \citet{\ReinersVI} suggest that an alternative process might be at work that is responsible for the strong shear observed in the first group.

An upper envelope can be identified to the horizontal shear of the second group while the four established differential rotators at the granulation boundary remain clearly above. \citet{\ReinersVI} identifies an upper envelope to the cooler F-type stars which rises between periods of 0.5 and 3 days and then declines towards larger rotational periods at constant $\alpha$. \citet{\ReinersVI} point out that the rising part of the upper bound to the horizontal shear of the second group between periods of 0.5 and 3 days could also be described by a plateau at $\Delta\Omega\approx0.7\,\radpd$. Fig.~\ref{fig:shear_P} strengthens this interpretation and shows an updated version of the upper bound of \citet[fig.~4]{\ReinersVI} suggesting a bit higher plateau close to values of $\Delta\Omega\approx1\,\radpd$. 
The envelope is transformed to the $\alpha$ scale (Fig.~\ref{fig:alpha_P}). The shear of all the F stars in the second group is below $1\,\radpd$ at periods of up to 3\,days while for longer periods, the shear is below the curve of maximum $\alpha=1$. 

It should be noted here that values of inclination very different from $90\deg$ may allow for shear measurements greater than 1. However, the measured shear scales with $\sqrt\sini$ and the probability to find an object with $\sini$ very different from 1 is small.

There are three particular objects, HD\,17094, HD\,44892, and HD\,182640 that would be located at the upper envelope in Fig.~\ref{fig:shear_P}. HD\,17094 and HD\,182640, however, are possibly affected by multiplicity while gravitational darkening might be important in the case of HD\,44892. Therefore, these objects are not considered.

In contrast to the F stars, the shear of the stars at the granulation boundary at the transition from spectral type A to F shows a clear period dependence -- though based on a few data points only -- and seems to line up roughly parallel to the curve $\alpha=1$. It cannot be excluded, however, that the line profiles of the hot stars with short rotational periods of less than a day are rather due to gravitational darkening, in parts at least, than due to differential rotation.

\subsection{Comparison to theoretical predictions}
Figures~\ref{fig:alpha_P} and \ref{fig:shear_P} compare the measurements to models of \citet{\KuekerV} who modeled differential rotation for an F8 and a G2 star and later for hotter F stars \citep{\KuekerVII}. They describe a strong dependence on rotational period. \citet{\KuekerV} notice maxima in the period dependence of shear in the F8 and G2 models. The maximum of the F8 star is higher and at shorter rotational periods. They further point out that the increase of maximum horizontal shear with increasing effective temperature is much stronger than the variation due to rotational period.

The comparison to the predictions of \citet{\KuekerV} is complicated since these cover longer periods of 4\,d and more. There is only a small overlap with the present work between 4 and 10 days. In this range, the measurements of late-F type stars scatter widely and are larger than the prediction for the F8 star. Only HD\,114642 agrees nicely with the prediction for an F8 star although the rotational period of this particular object is not much different from the other late-F differential rotators. None of the stars studied displays rotational shear below the shear predicted for an F8 star which is in agreement with the predictions as none of the differential rotators identified has spectral type G or later. A cool, slowly-rotating G-type star like the Sun (which is also highlighted in Figs.~\ref{fig:alpha_P} and \ref{fig:shear_P}) is beyond the detection limit.

Although the observed $\alpha$ is larger than predicted at spectral type late-F, the total range of relative differential rotation $\alpha$ observed roughly agrees with the total range predicted by \citet{\KuekerV}. A close inspection of Figs.~\ref{fig:alpha_P} and \ref{fig:shear_P} shows that it is not necessarily due to underestimated shear that the observed shear seems higher. Instead, the apparent underestimation might also be due to disagreements in period. The observed $\alpha$ occur at much shorter rotational periods than the predicted data which translates to higher values of shear. The consequence is that the measured shear is higher than predicted shear. In principle, there is a large uncertainty in observed periods because of unknown inclination. However, inclinations very different from $90\deg$ will result in even shorter periods as is indicated by the straight solid lines in Figs.~\ref{fig:alpha_P} and \ref{fig:shear_P}.

While a clear trend with rotational period is not discernible among the observations of the F stars, the upper envelope to $\alpha$, at least, partly has a shape similar to that of the curves predicted by theory. It can be noticed from Fig.~\ref{fig:alpha_P}, that the rising parts of these curves roughly follow the lines of constant horizontal shear.

The situation looks different for hotter F stars. Among later-type stars, \citet{\KuekerVII} also modeled F stars up to $1.4\,\Msun$ corresponding to spectral type F5 on the main sequence. A total range in horizontal shear is predicted (not shown in the figures) which roughly agrees with the horizontal shear observed. Again, a strong variation will be involved due to the dependence on rotational period. The rotational periods assumed in the calculations are not available to us so that we cannot study the behaviour of relative differential rotation $\alpha$. The theoretical calculations agree with observations in that horizontal shear of $1.0\,\radpd$ is not exceeded. So far, there are no predictions from models of more massive, rapidly rotating which would be very interesting however.

There are basic principles, both physical and conceptual, which strongly constrain observable shear. According to \citet{\KuekerVII}, as early F-type stars have thin convective envelopes, strong horizontal shear of the order of $1\,\radpd$ can only be sustained at rotational periods of the order of 1 day. This means that rotation necessarily has to be fast in order to sustain strong shear. On the other hand shear that strong cannot appear at slow rotation. Fig.~\ref{fig:shear_P} shows that only at short periods, relative differential rotation will not be above $100\,\%$. In other words, the relative differential rotation of a slower rotator with similarly strong shear would be larger than $100\,\%$.

\subsection{Horizontal shear and effective temperature}

\begin{figure}
\includegraphics[width=\figwidth]{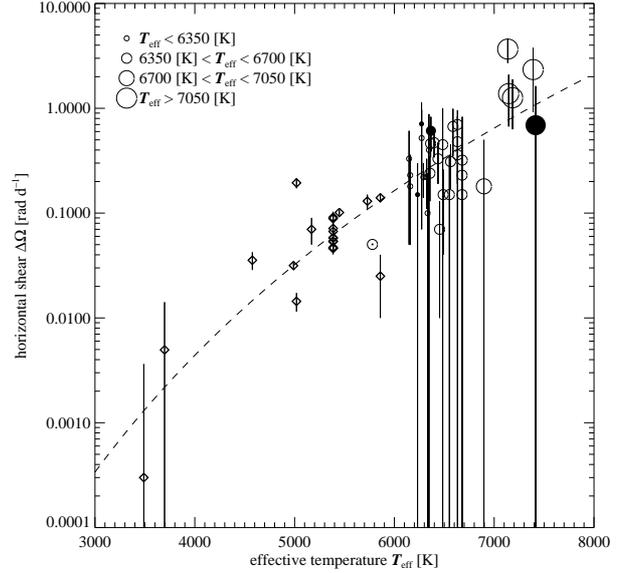}
\caption{\label{fig:shear_teff}The figure displays the absolute shear $\Delta\Omega$ vs. effective temperature for all measurements of differential rotation including previous measurements by other methods \citep[diamonds,][]{\BarnesV}, compared to the location of the Sun ($\odot$). The dashed line represents the fit by \citet{\BarnesV}. Circles identify results from line profile analysis as given in the caption to Fig.~\ref{fig:alpha_P}. Values derived from the $\alphasini$ calibration instead of $\alphaimax$ or $\alphaimin$ are presented in order to be consistent with previously published versions of this figure. Error bars due to unknown inclination are omitted for clarity.} 
\end{figure}
Figure~\ref{fig:shear_teff} updates fig.~5 of \citet{\ReinersVI} and displays the relation of horizontal shear and effective temperature. The measurements of the present work are added.

The trend detected by \citet{\BarnesV} is generally followed by the results from line profile analysis which fill the plot at spectral types F and earlier although the hotter differential rotators clearly stand out and display stronger shear. At spectral type F8 corresponding to an effective temperature of $\approx6200\,$K, horizontal shear covers a wide range of $\approx0.1-1.0\,\mathrm{d}^{-1}$ while the stars at the granulation boundary are above that and reach $\Delta\Omega\approx3\radpd$.

\subsection{Implications for the frequency of differential rotators}

An interesting question is how far the attainable values of horizontal shear and the rotational periods are related to the frequency of differential rotation seen across the HR diagram. The detectable values of $\alpha$ play a crucial role. Assuming the same shear $\Delta\Omega$ at different rotational periods, $\alpha$ will be lower at shorter periods and no longer detectable below a certain period that depends on shear. Therefore, stars with less shear will be more easily detected at slow rotation while they might be undetected at fast rotation \citep{\KuekerV,\ReinersVI,\KuekerVII}. As temperature and rotation are correlated, this might explain the general trend of less detections of differential rotation at higher effective temperature and faster rotation.

\begin{table}[!ht]
\caption{\label{tab:shear}Number of stars with measured rotational profile, distinguished by effective temperature and by rotational period.}
\begin{center}
\begin{tabular}{llccc}
\hline\hline
\noalign{\smallskip}
&&early-F&mid-F&\\
&&$7050-6700$\,K&$6700-6350$\,K&\\
\noalign{\smallskip}
\hline
\noalign{\smallskip}
P$<1$\,d &total&26&9&\\
&diff. rot.&0&2&\\
\noalign{\smallskip}
\hline
\noalign{\smallskip}
P$>1$\,d&total&48&55&\\
&diff. rot&1&15&\\
\noalign{\smallskip}
\hline
\end{tabular}
\end{center}
\end{table}

Together with the upper bound to horizontal shear of F stars, this might also explain the lack of differential rotators among early-F type stars. Adopting the upper bound to horizontal shear of F stars of $\Delta\Omega\approx1\,\radpd$ in Fig.~\ref{fig:shear_P} and a detection limit of $\alpha=0.05$, we can only detect differential rotation in stars rotating slower than $\Omega=\frac{\Delta\Omega}{\alpha}\approx\frac{1\,\radpd}{0.05}$, i.e. periods longer than $P=\frac{2\pi}{\Omega}\approx0.31$ days. This means that no differentially rotating F star is expected at periods much less than a day. This limit cannot be accessed by the measurements since the shortest detected period among the stars with rotational profile measured is $\approx0.4\,$d. It is instructive to look more closely at stars with $6700<\Teff\leq7050\,$K corresponding to early-F spectral types and compare them to stars with $6350<\Teff\leq6700\,$K in the mid-F range. Furthermore, we distinguish in both groups stars with rotational periods shorter than a day and stars with longer period (Table~\ref{tab:shear}). First of all, the numbers in the table agree with the finding that hotter stars tend to rotate faster. In other words, the fraction of early-F type stars with rotational periods less than 1\,d is larger than the fraction of mid-F type stars at the same periods.

Indeed there is no single rapid differential rotator with a rotational period of less than a day among the early-F stars. However, there are two differential rotators among the mid-F stars, approximately one fourth. Similarly, there are 15 differentially rotating mid-F stars at periods larger than 1 day, again one fourth, but there is only one early-F star. While this agrees with the previous finding that the frequency of differential rotators decreases towards hotter stars, it is somewhat unexpected that the fractions are the same for both period bins. This indicates that the rotational periods and the implied detection limit of $\alpha$ do not fully explain the gap of differential rotation among early-F stars.

In a certain respect, the lack of differential rotation among early-F stars contrasts the results from modeling. According to models by \citet{\KuekerVII}, horizontal shear will be higher at hotter effective temperature that would facilitate the detection of differential rotation. Therefore, a substantial fraction of differential rotators must be expected for early-F type stars even though they rotate faster than mid-F stars and even though, consequently, the detection limit on shear is higher. This is not the case however. A possible explanation might be found in the details of the period dependence and in particular the exact rotational periods at the predicted maximum of horizontal shear. For example, there will be less detections of differential rotation among early-F stars if the corresponding maximum is below $1\,\radpd$ and located at periods shorter than 1 day and if at the same time the maximum for mid-F stars is at longer periods. More modeling of early-F stars at very short periods is clearly needed.

\begin{figure}
\includegraphics[width=\figwidth]{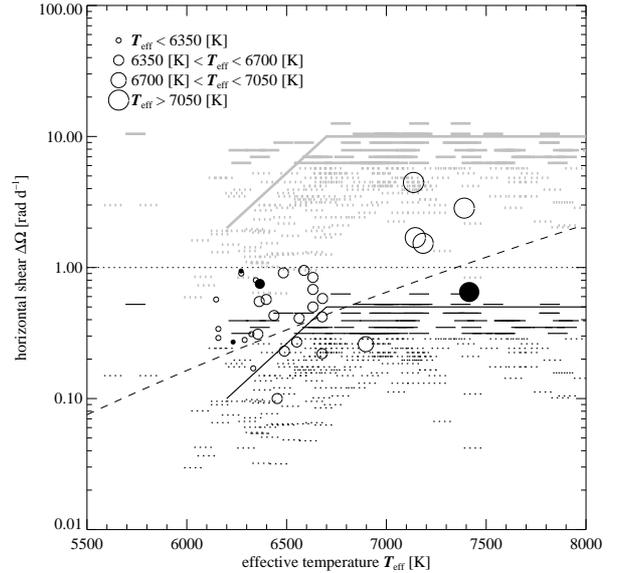}
\caption{\label{fig:shear_teff_lim} The figure enlarges the hot portion of Fig.~\ref{fig:shear_teff} and displays the absolute shear $\Delta\Omega$ vs. effective temperature for measurements by line profile analysis. Symbols identify measurements of differential rotation from line profile analysis as given in the caption to Fig.~\ref{fig:alpha_P}, but here assuming an inclination of 90$\deg$. In addition, the ranges of shear accessible to observation are indicated by short horizontal lines. The cases of $\alpha=0.05$ and $\alpha=1$ are distinguished by black and grey, respectively. These ranges are determined by rotational period alone, in a way that is given in Fig.~\ref{fig:shear_P}. Therefore, the detectable ranges can not only be given for those objects which display differential rotation. Instead they can be assessed for all objects with the rotational profile measured, including the stars that are considered rigid rotators. Among these limits indicated by short horizontal lines, stars with periods longer than a day are distinguished from stars with shorter periods by dotted lines. The two solid (black and grey) flexed lines represent envelopes to the bulk of the limits. These are based on the maximum $\vsini$ observed which is most probably due to $\sini{\lesssim}1$. Error bars are omitted for clarity. The dotted line denotes the upper bound on shear of F stars identified in Fig~\ref{fig:shear_P}. The dashed line reproduces the fit of \citet{\BarnesV}.} 
\end{figure}

Figure~\ref{fig:shear_teff_lim} enlarges the hot portion of Fig.~\ref{fig:shear_teff} and shows the temperature effect more clearly. The range of shear accessible to line profile analysis is shown for all stars with rotational profiles measured. The lower limits to measurable shear are derived from the rotational period determined for each star assuming $\alpha=0.05$. The upper limits are inferred the same way assuming $\alpha=1.0$. Therefore, these two limits enclose the whole range of measurable shear for each star. In accordance to Fig.~\ref{fig:shear_P}, highest values of shear are accessed at very short rotational periods $P<1\,$d.

Error bars due to inclination are omitted. Lower values of inclination will essentially increase the shear measurements but also raise the limits and the upper bound $\Delta\Omega=1\,\radpd$ identified in Fig.~\ref{fig:shear_P}.

The distribution of the limits with respect to effective temperature actually reflects the correlation with $\vsini$ as periods were calculated from $\vsini$ and photometric radii. The detection limits generally decline towards cooler effective temperatures since rotational periods increase and thus smaller amounts of shear are detectable. At the same time, extremely large values like for the transition objects at the granulation boundary cannot be detected any more.

A major conclusion from Fig.~\ref{fig:shear_teff_lim} is that the fit of \citet{\BarnesV} in Fig.~\ref{fig:shear_teff} cannot be tested with our data. The detectable range of shear is constrained by selection effects. The lower bound results from the detection limit on $\alpha(=0.05)$ while the upper bound is given by $\alpha=1$. The correlation of faster rotation with higher $\Teff$ explains at least in parts the trend of increasing shear with hotter effective temperature.

The case of high inclination deserves closer inspection since Fig.~\ref{fig:shear_teff_lim} shows values of shear assuming an inclination of $90\deg$ to facilitate comparison with Fig.~\ref{fig:shear_P}. It is a reasonable assumption that the stars with the highest $\vsini$ measured will be those with $\sini{\lesssim}1$. Consistently, these stars will have the shortest rotational periods so that the upper bound to $\vsini$ corresponds to a lower bound to periods. Shear is derived from periods and from $\alpha$ measured by line profile analysis such that a lower bound to periods corresponds to an upper envelope to the limits on shear. The upper bounds to the bulk of the limits are indicated by the blue and red flexed lines in Fig.~\ref{fig:shear_teff_lim}.

These upper bounds consist of a rising part at temperatures of mid-F spectral types and by a flat part at hotter temperatures. The bound to the lower limit of detectable shear (black solid line) and the upper bound to the shear of F stars of $1\,\radpd$ identified in Sect.~\ref{sect:shear_period} (dotted line) define a detectable range of shear which becomes very narrow towards hotter temperatures ($\approx0.5-1\,\radpd$ at $\Teff\approx6600\,$K). Therefore, differentially rotating F stars with horizontal shear below $0.5\,\radpd$ will not be detected if their projected rotational period is shorter than a day. However, F stars with shear higher than $\approx0.5\,\radpd$ should still be detectable even at rotational periods that short. Furthermore, the number of lower and upper limits in Fig.~\ref{fig:shear_teff_lim} at effective temperatures above $6600\,$K is large so that there are many suitable objects at early-F spectral types. However, all these seem to be rigid rotators or, at least, display undetectable rates of differential rotation. Therefore, the detection limits to differential rotation do not fully explain the lack of differential rotators at early-F spectral types.

In summary, three issues have been identified which play an important role in explaining the lack of differential rotators among early-F type stars: an apparent upper bound to horizontal shear of F-type stars, the detection limit on shear which depends on rotational period, and the correlation of rotational periods and effective temperature. However, these cannot fully explain the lack of differential rotators among the early-F type stars. A direct dependence on temperature is apparent. We speculate that the physical conditions and/or mechanisms of differential rotation change at early-F spectral types. This change might also be related to the detection of extremely large horizontal shear at the transition to A spectral types.

\section{Summary and outlook}

In the present work, stellar rotation of some 180 stars has been studied through the analysis of the Fourier transform of the overall spectral line broadening profile, extending previous work of Reiners {\etal} 2003-2006. In total, the rotation of 300 stars has been studied by line profile analysis. Among these, 33 differential rotators have been identified.

In order to study frequency and strength of differential rotation throughout the HR diagram, particular emphasis was given to the compilation of astrometric and photometric data which has been drawn homogeneously from the Hipparcos and Tycho-2 catalogues. 
Complications introduced by multiplicity were discussed in detail and affected measurements flagged. Multiplicity might explain exceptional amounts of differential rotation, e.g. in the case of HD\,64185\,A. The flagged objects were not included in the discussion. It will be worthwhile to further study these objects since they are are located in interesting regions of the HR diagram.

The results were found to be in good agreement with previous observations for stars in common. Substantial deviations can be explained by instrumental limitations or peculiar line profiles due to binarity or spots. Otherwise discrepancies exceed 5\,$\kmps$ in $\vsini$ and 0.10 in $\qfrac$ only when $\vsini$ is below the limits for line profile analysis (e.g. $12\,\kmps$ for CES and $45\,\kmps$ for FEROS).
A compilation of all stars with new or previous rotation measurements by line profile analysis was presented.

The present work detected several rigid rotators among A stars. Some rapidly rotating A stars show line profiles probably affected by gravitational darkening. As of now, no differential rotators are known with effective temperatures higher than 7400\,K.

The frequency of detections of differential rotators generally decreases with increasing effective temperature and increasing projected rotational velocity. Effective temperature and rotational velocity are correlated because of the dependence of magnetic braking on spectral type. This stops us from identifying the quantity responsible - temperature or rotation.

On the other hand, the fraction of differential rotators does not vary notably with surface gravity. In the present work, surface gravity was estimated from photometry to be used as indicator of the luminosity class. The study of the rotational behavior of evolved stars will certainly benefit from a more accurate assessment of surface gravity and allow for a comparison with theoretical predictions \citep[e.g.][]{1999A&A...344..911K}. The analysis of giants will be particularly promising since at the same $\vsini$ longer rotation periods can be studied. Of course, the challenge is to find giants with sufficiently large $\vsini$ to be accessible to line profile analysis.

The lack of differential rotators around $T_\mathrm{eff}=6750\,$K (spectral type F2-F5) detected in previous work was consolidated. At the cool side of this gap, there is a large population of differential rotators. The differential rotator with the highest differential rotation parameter measured so far by line profile analysis  (HD\,104731, $\alphasini=54$\,\%) is right at the cool edge of this gap. This object and also other strong differential rotators in this region exhibit moderate rotational velocity. At the hot side of the gap at the location of the granulation boundary, all differential rotators are rapid rotators ($\vsini\gtrsim100\,\kmps$) and show strongest horizontal shear ($\Delta\Omega\gtrsim1\,\radpd$). 


Theoretical calculations are at hand for F-stars and were compared to the observations. While the observed range of horizontal shear in mid-F type stars agrees with the predicted range, the observed values in late-F type stars were found to be much larger than predicted. This was explained rather by a discrepancy in period dependence than in absolute shear strength.  Further observations with other techniques are needed to measure differential rotation at periods of 20\,d and more. Strong shear seems possible for very rapid rotators only. Models of more massive and rapidly rotating stars are clearly needed.

Even more, it was found that the observations of F stars do not follow a clear period dependence. Still, the upper envelope which traces the upper bound to horizontal shear and the curve of relative differential rotation $\alpha=1$ exhibit a shape similar to the predicted period dependence of an F8 and a G2 star.

Regarding the comparison with theoretical predictions, it is worth discussing the role of inclination. In the present work, the uncertainties of inclination were properly accounted for by deriving relative differential rotation and horizontal shear assuming different, extreme values of inclination. A remarkable fact is that a better assessment of inclination will not allow for a better agreement with the predicted period dependence. Instead, limb darkening and the assumed surface rotation law might be more critical sources of uncertainties in this respect. This does not mean that there is no need to better assess inclination, as it causes large part of the uncertainty about measurements of differential rotation.

 The horizontal shear of F stars remains below a plateau of $\Delta\Omega\approx1\,\radpd$. This might explain the low frequency of differential rotation among hotter and rapidly rotating F stars considering the detection limit on relative differential rotation $\alpha$ but cannot fully account for the gap of differential rotation at early-F spectral types. We assume, that conditions change in early-F type stars so that differential rotation is inhibited while at somewhat hotter temperatures at the transition to spectral type A, differential rotation reappears with extreme rates of differential rotation.

The shear of rapidly rotating A stars was found to be higher than $1\,\radpd$ and possibly increases with decreasing rotational period. So far, there are no predictions from models which could explain the shear of A-type stars. Although \citet{\KuekerVII} set out to understand the strong shear detected at the granulation boundary by \citet{\ReinersIV}, the stellar models discussed are not beyond 1.4\,$\Msun$, i.e. mid-F type stars.

The strength of differential rotation generally follows the increasing trend with effective temperature identified by \citet{\BarnesV} but the trend with effective temperature can actually not be tested with our data since the spread is large and the detection limits on shear also follow the trend with temperature as $\vsini$ and $\Teff$ are correlated.

As the surface rotation law is identified as a possible source of uncertainty, it is important to emphasize that the present work is based on the assumption of a very simple surface rotation law (Eq.~\ref{eq:rotlaw}). This assumption is reasonable since it is motivated by the solar case and has been consistently used by similar studies. However, \citet{2008JPhCS.118a2029K} point out that in the case of the Sun there are other measurements favouring a rotation law dominated by a ${\cos}^4$ term of co-latitude instead of the ${\cos}^2$ term \citep{1984SoPh...94...13S}. The adoption of such a rotation law will introduce a second parameter of differential rotation and might change the shear measurements from line profile analysis.

\begin{acknowledgements}
We would like to thank our referee, Dr. Pascal Petit, for a very clear and constructive report. M.A. thanks Dr. Theo Pribulla for discussion which helped to improve the analysis of the A stars among the sample. M.A. and A.R. acknowledge research funding granted by the {\it Deutsche Forschungsgemeinschaft} (DFG) under the project RE 1664/4-1. M.A. further acknowledges support by DLR under the projects 50OO1007 and 50OW0204. This research has made use of the {\it extract stellar} request type of the Vienna Atomic Line Database (VALD), the SIMBAD database, operated at CDS, Strasbourg, France, and NASA's Astrophysics Data System Bibliographic Services.
\end{acknowledgements}

\bibliographystyle{aa}
\bibliography{diffrot}

\Online

\end{document}